\documentclass[preprint]{aastex}
\begin{document}

\newcommand{\Msol}{M$_{\odot}$}
\newcommand{\kms}{km s$^{-1}$}
\newcommand{\degsq}{deg$^2$}
\newcommand{\perdegsq}{deg$^{-2}$}
\newcommand{\mgb}{Mg~$b$}
\newcommand{\by}{$b\!-\!y$}
\newcommand{\mi}{$M\!-\!T_2$}
\newcommand{\mg}{$M\!-\!51$}
\newcommand{\cm}{$C\!-\!M$}
\newcommand{\vi}{$V\!-\!I$}
\newcommand{\ttwo}{$T_2$}

\title{Mapping the Galactic Halo I. The ``Spaghetti'' Survey}
\shorttitle{Mapping the Halo}

\author{Heather L. Morrison\altaffilmark{1,2}}
\altaffiltext{1}{Cottrell Scholar of Research
Corporation and NSF CAREER fellow}
\altaffiltext{2}{and Department of Physics}
\affil{Department of Astronomy,
Case Western Reserve University, Cleveland OH 44106-7215 
\\ electronic mail: heather@vegemite.astr.cwru.edu}

\author{Mario Mateo}
\affil{Department of Astronomy, University of Michigan, 821
Dennison Bldg., Ann Arbor, MI 48109--1090\\
electronic mail: mateo@astro.lsa.umich.edu} 
\author{Edward W. Olszewski}
\affil{Steward Observatory, University of Arizona, Tucson,
AZ 85721\\
electronic mail: edo@as.arizona.edu} 
\author{Paul Harding}
\affil{Steward Observatory, University of Arizona, Tucson, Arizona 85726
\\ electronic mail: harding@billabong.astr.cwru.edu}
\author{R.C. Dohm-Palmer}
\affil{Department of Astronomy, University of Michigan, 821
Dennison Bldg., Ann Arbor, MI 48109--1090\\
electronic mail: rdpalmer@astro.lsa.umich.edu}
\author{Kenneth C. Freeman}
\affil{Mount Stromlo and Siding Spring Observatories, ANU, Private Bag,
Weston Creek PO, 2611 Canberra, ACT, Australia\\
electronic mail: kcf@mso.anu.edu.au}
\newpage
\author{John E. Norris}
\affil{Mount Stromlo and Siding Spring Observatories, ANU, Private Bag,
Weston Creek PO, 2611 Canberra, ACT, Australia\\
electronic mail: jen@mso.anu.edu.au}
\and
\author{Miwa Morita}
\affil{Steward Observatory, University of Arizona, Tucson,
AZ 85721\\
electronic mail: mmorita@as.arizona.edu} 


\begin{abstract}

We describe a major survey of the Milky Way halo designed to test for
kinematic substructure caused by destruction of accreted
satellites. We use the Washington photometric system to identify halo
stars efficiently for spectroscopic followup. Tracers include halo
giants (detectable out to more than 100 kpc), blue horizontal branch
stars, halo stars near the main sequence turnoff, and the ``blue
metal-poor stars'' of \citet{pbs94}. We demonstrate the success of our
survey by showing spectra of stars we have identified in all these
categories, including giants as distant as 75 kpc. We discuss the
problem of identifying the most distant halo giants.  In particular,
extremely metal-poor halo K dwarfs are present in approximately equal
numbers to the distant giants for $V>18$, and we show that our method
will distinguish reliably between these two groups of metal-poor
stars.

We plan to survey 100 square degrees at high galactic latitude, and
expect to increase the numbers of known halo giants, BHB stars and
turnoff stars by more than an order of magnitude.  In addition to the
strong test that this large sample will provide for the question ``was
the Milky Way halo accreted from satellite galaxies?'', we will
improve the accuracy of mass measurements of the Milky Way beyond 50
kpc via the kinematics of the many distant giants and BHB stars we
will find.

We show that one of our first datasets constrains the halo density law
over galactocentric radii of 5--20 kpc and z heights of 2--15 kpc. The
data support a flattened power-law halo with b/a of 0.6 and exponent
--3.0. More complex models with a varying axial ratio may be needed
with a larger dataset.

\end{abstract}


\section{INTRODUCTION}
How much of the Galaxy's halo was accreted from satellite galaxies?
What fraction of these accretions have left
substructure that we can detect today?
Hierarchial galaxy formation pictures \citep{davis85,gov97,klyp99}
suggest that structure forms first in small clumps which later combine
to make larger galaxies. While this picture describes dark matter
rather than stars, it is reasonable to expect that some stars would
have formed in these dense clumps of matter at early times. This is
borne out by the fact that almost all of the Local Group dwarf
galaxies contain stars with ages greater than 10 Gyr \citep{mm98}.
In studies of the Milky Way, the first suggestion that the Galaxy's
halo did not form in a fast, uniform collapse \citep{egg62} was made by
\citet{szinn}, who noted that the horizontal-branch morphology of the
outer halo globulars could be explained by a younger mean age. These
clusters would have originated 
in ``transient protogalactic fragments that continued to
fall into ...... the Galaxy for some time after the collapse of its
central regions''.  

Strong variations in the Galaxy's potential associated with the
formation of the inner disk and bar, plus the shorter orbital
timescales there, may have erased the kinematic signature of halo
accretion in its inner regions. Substructure may persist for many Gyr
further from the galactic center \citep{jsh,paul99}. We see evidence
for accretion not only in the Sgr dwarf galaxy, which is being
tidally disrupted on its current passage close to the Galaxy's disk \citep{igi}, but also in the
detection of various moving groups in the halo field
\citep{srm94,helmi}. These latter objects are particularly interesting
and surprising because they
are at relatively small distances from the galactic center (8--10 kpc).

\begin{figure}
\includegraphics{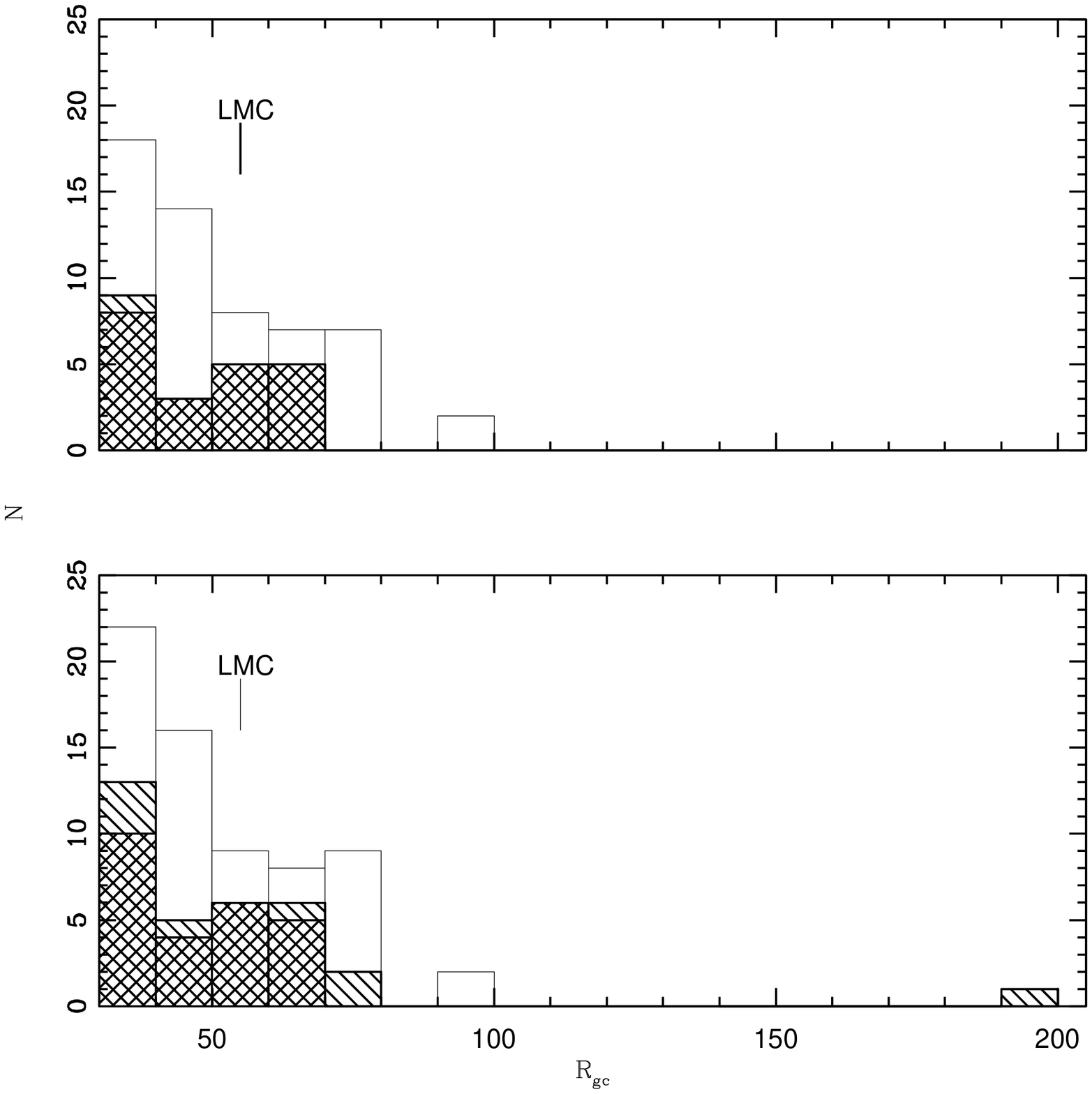}
\figcaption[Morrison.f1.eps]{(a) Histogram of galactocentric distances to halo field stars
with $R_{gc} > $ 30 kpc known to date. Horizontal branch stars (RR
Lyrae variables and BHB stars) are shown cross-hatched, red giants
shaded and carbon stars comprise the rest. References for sources are
given in Table \ref{distrefs}. (b) Histogram of distances to outer
halo field stars with the red giants and BHB stars of \protect\citet{edo99}
added -- these stars were all confirmed spectroscopically during a
single KPNO 4m run. The giant at 190 kpc needs a confirming spectrum
with higher S/N before we can be 100\% confident of its
luminosity. \label{disthist}}
\end{figure}

Figure \ref{disthist}(a) shows in histogram form the numbers of
distant halo objects known to date. Globular clusters and dwarf
spheroidal galaxies are not included. It is not surprising, with such
small samples, that almost all the discoveries of halo substructure to
date have been serendipitous. In this paper, we will describe a
survey which will give a {\it quantitative} answer to the question
``how much of the halo was accreted?'' by identifying a sample
comprising a large number of halo stars out to great distances.
(Figure \ref{disthist}(b) shows how the situation has improved after
our first spectroscopic followup run on the KPNO 4m.)

In Section 2 we discuss the design of our survey and the tracers we
use, together with the region of the halo that each tracer will
sample.  Section 3 discusses our photometric selection technique in
detail for each tracer, showing its efficacy with spectra of
stars found in each category. We also discuss the possible
contaminants of our sample and how we reject them. Section 4 uses the
numbers of turnoff stars found by our survey in various directions to
constrain the density distribution of the halo. We also report the first
evidence of spatial substructure in the halo.

Future papers in this series will discuss our simulations of the
breakup of satellites and their observational consequences
\citep{paul99}, our photometric survey
\citep{robbie}, our spectroscopy of distant metal-poor giants and 
BHB stars \citep{edo99} and evidence for spatial substructure in our
photometric data \citep{hlm00}.

\section{SURVEY DESIGN}

Because substructure is visible in velocity space long after it
disappears in density space \citep{jsh,paul99}, we aim
to obtain velocity data on a large sample of halo stars.
\citet{paul99} discuss the signature that tidally disrupting streams
will show in velocity histograms: although the observed signature does
depend on the viewing geometry and initial conditions of the
accretion, in many cases velocity histograms will be bimodal or
multi-modal. The features that correspond to disrupted satellites
will show a velocity dispersion of order tens of kilometers per second.
Since
substructure will survive longest in the outer Galaxy, distant halo
tracers are advantageous. If we restrict our spectroscopic follow-up
to 4m-class telescopes, this means that halo red giants and blue
horizontal branch stars are the tracers of choice because of their
intrinsic luminosity. However, these stars are intrinsically rare, and
so they limit our detection methods for substructure --- for example,
it is impossible to find a sample of 100 halo giants in a field of
size 1 deg$^2$. Samples of this size are needed for methods of
substructure detection based on velocity histograms. In order to have
more sensitivity to subtle signatures of accretion, we also need
tracers which are more numerous, such as halo turnoff stars. We shall,
however, not be able to probe the extreme outer halo with these less
luminous stars until 8m-class telescopes are more generally available.

The stellar halo provides a very small fraction of the Milky Way's
luminosity --- in the solar neighborhood, the disk-to-halo star number
ratio is $\sim$800:1, \citep{bc86,hlm93}, and even the thick disk
outnumbers the halo by $\sim$40:1. Thus it is important to use a
method which will {\it efficiently} pre-select halo stars before
obtaining velocities.

\subsection{Sky Coverage}

When aiming to answer a question about the origin of the entire halo,
all-sky coverage would be ideal.  \citet{jhb}
discuss such a survey for halo substructure, which,  unfortunately,
is not feasible at present. Existing
all-sky surveys based on photographic plates cannot produce
photometry accurate enough to efficently select halo objects apart from
 rare blue horizontal branch stars.  Any spatial
substructure in the halo will be washed out by the many foreground
disk and thick disk stars in a photographic survey.
The Sloan survey \citep{gunn} plans to cover one quarter of the sky at
high galactic latitude in the North. The photometry from this survey will be
sufficiently accurate to enable the identification of halo turnoff
stars and BHB stars, but not the rare distant red halo giants we
discuss below. These stars, which
will be inseparably contaminated by foreground K dwarfs in the Sloan
colors, are particularly valuable for this project
because they probe the extreme outer halo.

We have chosen to use a pencil-beam survey of various
high-latitude fields, using CCD photometry and the Washington system
to select the halo stars, and then carry out follow-up spectroscopy 
to search for kinematic substructure.

Even with the most extreme hypothesis that the entire halo is
composed of tidal streams, their filling factor on the sky will still
be small. Some fields will therefore contain more stars than the average and
some considerably fewer. A large number of different pointings is
preferable to maximize the chance of hitting a single stream. For
example, assuming a typical dimension for a stream of 2 degrees by 100
degrees, 50 of these would cover only 25 percent of the sky. We have
chosen to survey initially 100 square degrees of the sky, in 100 different
fields randomly distributed at high galactic latitude (generally above
$|b|=45$), with most fields having galactic longitude between 90 and
270.

We chose in general to stay away from the quadrants which include the
galactic center for two reasons. First, structure is more readily
destroyed close to the galactic center, which lowers our chance of
detecting it. The best place to look for substructure is the outer
halo, where dynamical times are long and tidal forces small. Thus,
the galactic anticenter is better since the galactocentric radii of
the stars we detect will be larger. Second, there are many 
components represented at the galactic center -- young and old disk, thick disk
and bar, as well as the inner halo.   Interpretation of our
results would be more complex there due to the dynamical effect of the bar.
The metal-weak tail of the thick disk is also minimized when we
look at higher latitudes.

\subsection{Tracers}

Traditional ways of searching for halo stars include:
\begin{itemize}
\item Proper motion surveys
\citep{sf87,cl87,sean91}. These surveys identify stars
at most a few hundred pc away, except for the surveys of \citet{srm92}
and \citet{mend99},
which produced complete proper motion information on a sample of stars
to B=22.5 and 19 respectively (a maximum distance of 30 kpc).
\item RR Lyrae surveys \citep{kin65,abi,nick91,wett96,tdk96}. These
surveys sample more distant objects, but because of their extreme
rarity -- of order one per \degsq -- few distant RR Lyraes are known.
\item Objective prism surveys, generally for metal-poor giants or
stars near the main sequence turnoff.
\citep{bm73,rat89,bps85,mff}. Because of the relatively high
resolution needed to identify these stars spectroscopically, these
surveys are in general restricted to relatively bright magnitudes and
therefore relatively nearby objects.
\item Blue horizontal-branch (BHB) star surveys. The BHB stars are
identified either from their unusually blue color \citep{slc,nh91} or by using objective prism spectra near the Ca K
line \citep{pier82,pier84,bps85}. BHB stars are almost as rare as RR Lyrae variables, and
their identification is complicated by the presence of halo blue
stragglers, which have the same broadband colors as BHB stars but
higher gravity. As Norris and Hawkins show, the blue straggler
fraction may be as high as 50\% for samples of faint blue stars.
\item Carbon stars. These are extremely rare stars (of order one per
200 deg$^2$, Totten and Irwin 1998) which are identified easily from
objective-prism spectra. \citet{tot98} review the currently known halo
carbon stars and their properties. Distances are more uncertain for
these stars than for any other tracer we have discussed. 
\end {itemize}

Halo tracers we chose to use are: 
\begin{itemize}
\item Red giants. These are identified photometrically using a luminosity indicator based on the
\mgb/MgH feature at 5170\AA~~ plus a metallicity indicator based on
line blanketing near 4000\AA~, with spectroscopic confirmation.

\item Blue horizontal branch (BHB) stars, identified by their color
plus a spectroscopic check of gravity.
\item Stars at the main sequence
turnoff, identifiable from their blue color --- the most metal-poor
globular clusters have turnoff colors of \bv$\simeq$0.38, while the
thick disk turnoff is  \bv$\simeq$0.5, so stars with colors between
these two values are most likely  halo turnoff stars.
\item Blue metal-poor stars \citep{pbs94}, which are halo field
stars with colors bluer than \bv=0.38, thought to be younger main
sequence stars. When found in globular clusters, such stars are
typically referred to as blue stragglers and may have a different
origin from the field analogs.
\end{itemize}
We will discuss these tracers, and possible contaminants in our
sample, below.  We reject another halo tracer, RR Lyrae variables,
because of the large amount of telescope time needed to identify and
phase these variable stars.

Our technique reliably identifies both turnoff stars from the halo and
the more distant halo giants and BHB stars. Because of their different
luminosities, these objects probe different regions of the galactic
halo. Turnoff stars can only reach to galactocentric distances of
15--20 kpc using 4m-class telescopes for spectroscopic followup, while
red giants and BHB stars will reach to distances of more than 100
kpc. However, due both to shorter evolutionary timescales for giants
and to the strong decrease in halo density with galactocentric
distance ($\rho \propto r^{-3}$ or $r^{-3.5}$, Zinn 1985, Saha 1985),
there are far fewer halo giants detected than turnoff stars. Different
techniques are therefore used to search for the kinematic signatures
of accretion. For the more numerous turnoff stars, we use statistical
techniques based on the appearance of the velocity histogram (testing
for multimodality, for example, see \citet{paul99}). For the
rarer red giants, we use the technique of \citet{lb2} (see also
Lynden-Bell 1999) to search for stars with similar energies and
angular momenta, indicating a common origin.

\section{SELECTION TECHNIQUE}

Our initial survey was done using the Burrell and Curtis Schmidt
telescopes, which have CCD fields of order 1 deg$^2$.  The Burrell
Schmidt is fitted with a back-illuminated SITe 2048$\times$4096 CCD
with 1.5 arcsec per pixel, while the Curtis Schmidt (often) has a
2048$\times$2048 back-illuminated Tek CCD with 2.4 arcsec per
pixel. Now that large mosaics are available, we have extended the
survey using the CTIO 4m with the BTC (field = 0.25 deg$^2$) and the
8-CCD NOAO mosaics.  Our spectroscopic followup observations have been
made using the Hydra multiobject fiber spectrograph on the 3.5m WIYN
telescope and the RC spectrograph on the KPNO 4m. Future observations
with the Hydra spectrograph on the CTIO 4m and the Magellan telescope
are planned.

The Washington photometric
system combines strong metallicity sensitivity for late-type giants
with broad filter passbands, which contribute to observing
efficiency. We use this system for our survey, as its
filters can be used for selection of all the other tracers we need as
well. We describe transformations between the Washington system and
the BVI system in the Appendix.

In each survey field (of area approx 1 deg$^2$) we obtain 
photometry using a modified Washington \citep{can76,doug84} filter set (C,M,
51 and i' filters).  The large pixel area on
the Schmidt telescopes leads to a high sky level in the I band. Thus
we use the Sloan i' filter, whose passband avoids the worst of the
bright sky lines in the I band, in place of the Washington \ttwo\ or I filters. This
i' filter transforms readily to the Washington system.

Typical exposure times using back-illuminated CCDs and the 24-inch
Schmidt telescopes are 6000 sec in C and M, 8400 sec in 51, and 4800
sec in i'.  On the CTIO 4m with the BTC mosaic, exposure times were
100 sec in M, 500 sec in C, 120 sec in \ttwo\ and 250 in 51.  These give
typical errors of 0.015 mag. in each filter for a V=19 star.  Our
photometry will be discussed in more detail by \citet{robbie}.

We have used the data
of \cite{schlegel} to estimate the values of reddening in our fields,
and de-reddened the Washington colors according to the prescriptions
of \citet{can76} and \citet{hc79}. The reddening values are small, so
this did not have a strong effect on our results in any case.

\begin{figure}
\plotone{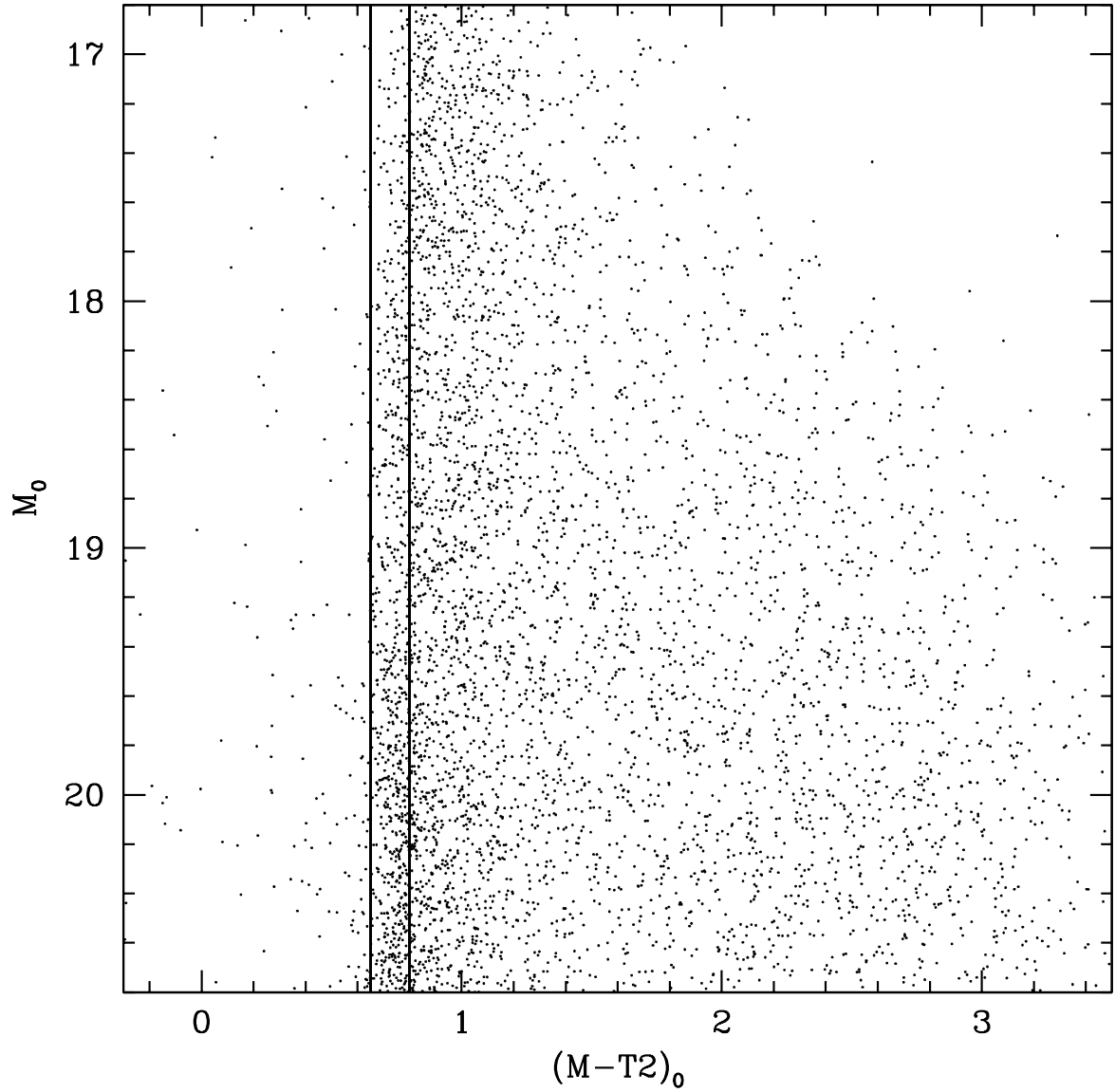}
\figcaption[Morrison.f2.eps]{Color-magnitude diagram (\mi\ vs $M$) for BTC
fields of area 2.75 deg$^2$. Halo and thick disk turnoffs are marked:
the halo turnoff is the bluer one, at $(M-T2)_0$=0.64.
  \label{cmd}} 
\end{figure}

Figure \ref{cmd} shows a typical color-magnitude diagram  with the
position of the halo and thick disk turnoffs marked.
Since each tracer requires a different selection technique, we will
discuss them separately.

\subsection{Halo giants}

This tracer is the most exciting since it  allows us to probe the
extreme outer halo. These stars have been little used in
the past because they are greatly outnumbered by foreground K dwarfs,
and it is difficult to distinguish K dwarfs from giants without
accurate intermediate-band photometry or spectroscopy. However, their potential is
enormous: a metal-poor star near the giant branch tip with $M_V$=--2
and $V$=19.5 (easily observable at medium resolution on a 4m-class
telescope) has a distance of 200 kpc!  The combination of large CCD
fields with the Washington photometric system makes the detection of
such objects feasible for the first time.

These distant halo giants are rare.  Using the simple model of
\citet{hlm93}, we find that there are of order 1--10 halo giants per
square degree down to V=20, using a range of assumptions about the
halo density distribution.  In section 4 we will show that out to
galactocentric distances of 20 kpc, the halo density law is well
described by a flattened power-law with exponent --3.0 and flattening
b/a=0.6. If this density law continues to larger distances, we would
expect to see 4 halo giants \perdegsq\ brighter than V=20 at the NGP
and 5 \perdegsq\ in an
anticenter field with galactic latitude 45$^\circ$.

There are three classes of objects found in the same range of color
and magnitude as the halo giants we wish to find:
\begin{itemize}
\item numerous K dwarfs of the thin and thick disk, which can be
detected using a photometric survey (\mg\ color)
\item extremely metal-poor halo dwarfs, which are present in
comparable numbers to halo giants for $V>18$, and need a good follow-up spectrum
(with S/N$\sim$15) to distinguish from halo giants.
\item background objects such as compact galaxies and QSOs, which are
easily weeded out using a low S/N followup  spectrum.
\end{itemize}

\subsubsection{Disk Dwarfs}

The major source of contamination of a halo giant sample is foreground
K dwarfs.  To quantify the numbers of foreground dwarfs that we will
need to weed out, we have estimated their number using a modified version of the
Bahcall-Soniera model which includes both thin disk and thick disk,
using a 5\% normalization for the thick disk.  This predicts 70 thin
disk dwarfs and 90 thick
disk dwarfs  ($0.9 <$ \bv$<1.2$) per deg$^2$ at the
NGP. 

Classical spectroscopic luminosity indicators (originally developed
for Pop. I stars, eg Seitter 1970) which are useful for our survey
include: 
\begin{itemize}
\item the \mgb\ triplet and MgH band near 5200\AA. These features are
much stronger in dwarfs than giants in this color range. They begin to
lose sensitivity to luminosity blueward of \bv\ = 0.9 (\mi=1.2). These
features are also temperature sensitive.
\item the Ca I resonance line at 4227\AA~ shows marked sensitivity to
both luminosity and temperature in K stars. Its strength increases as
temperature and luminosity decrease. Since we have an independent
photometric measure of temperature from \mi\ color, we can use the
Ca I 4227\AA~ line as a luminosity indicator. 
\item The blue and UV CN bands (bandheads at 4216 and 3883\AA~) are
strong in giants and not in dwarfs. The bands become weaker with
decreasing metallicity, and are not visible below [Fe/H] = --1.5.                                                                             
\end{itemize}

We have observed metal-poor dwarfs and giants in order to check whether
these indicators retain their usefulness for more metal-poor stars.
Our observations, plus those of the metal-weak giants we found, will
be discussed in more detail by \citet{edo99}.
We use the \mgb/MgH region as our major method of rejecting disk dwarfs
in our sample via photometric selection.  

\begin{figure}
\includegraphics[scale=0.8]{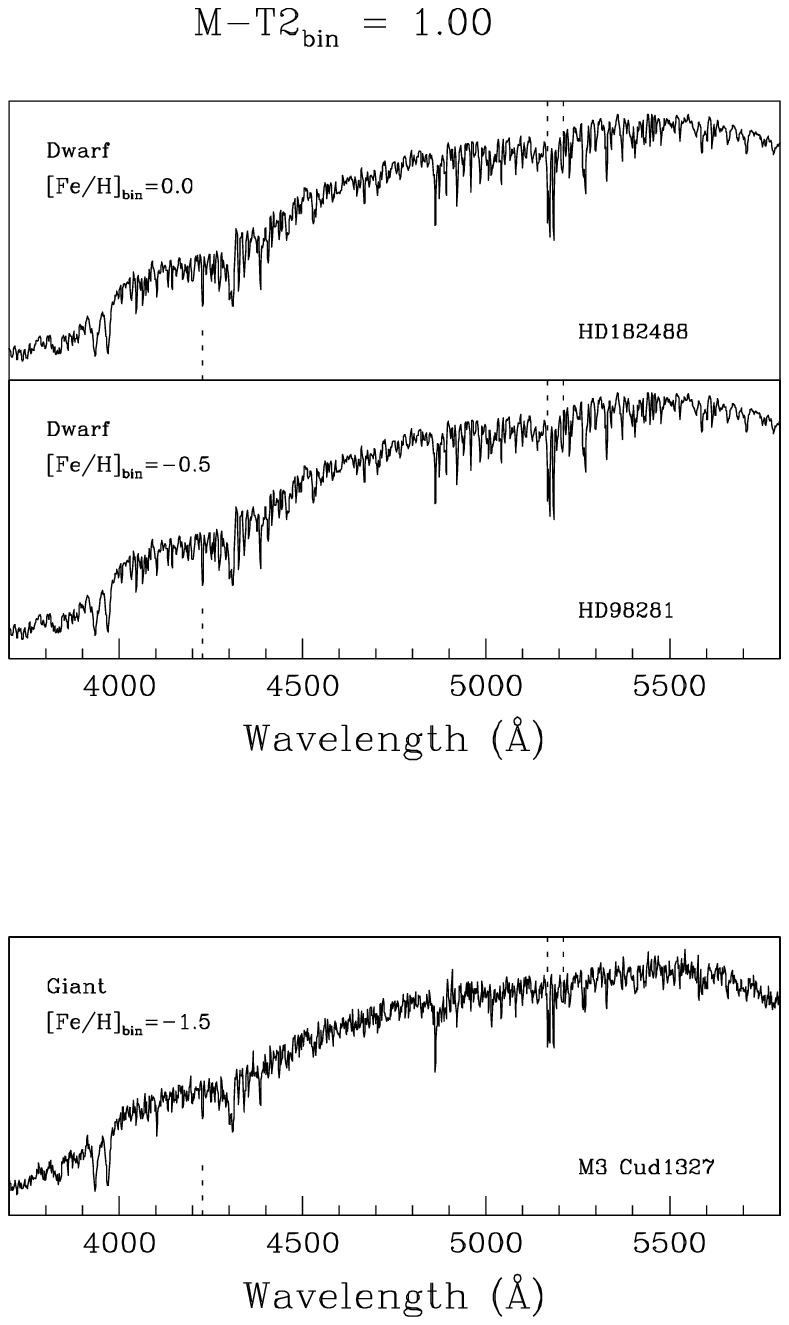}
\figcaption[Morrison.f3a.eps,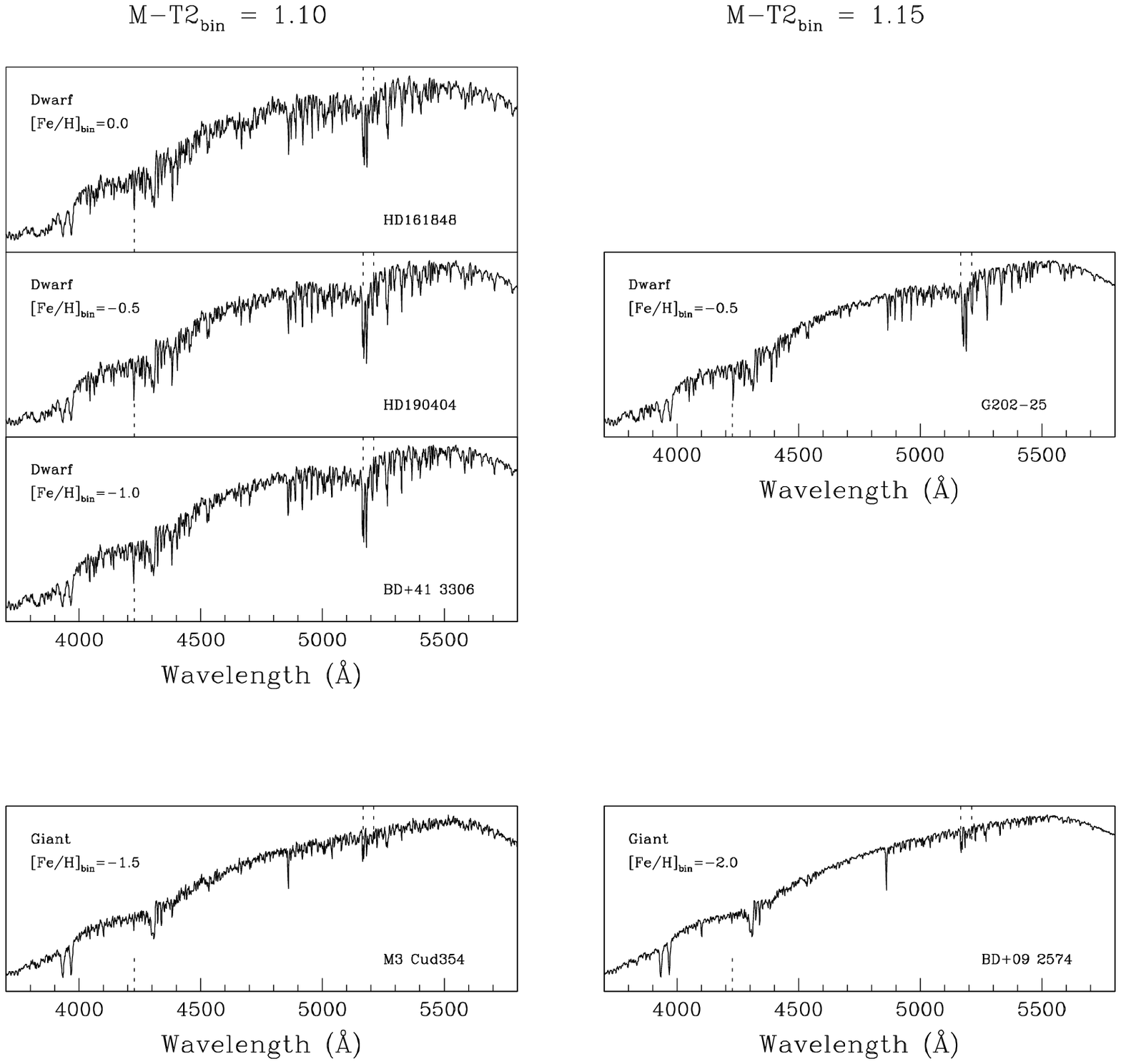,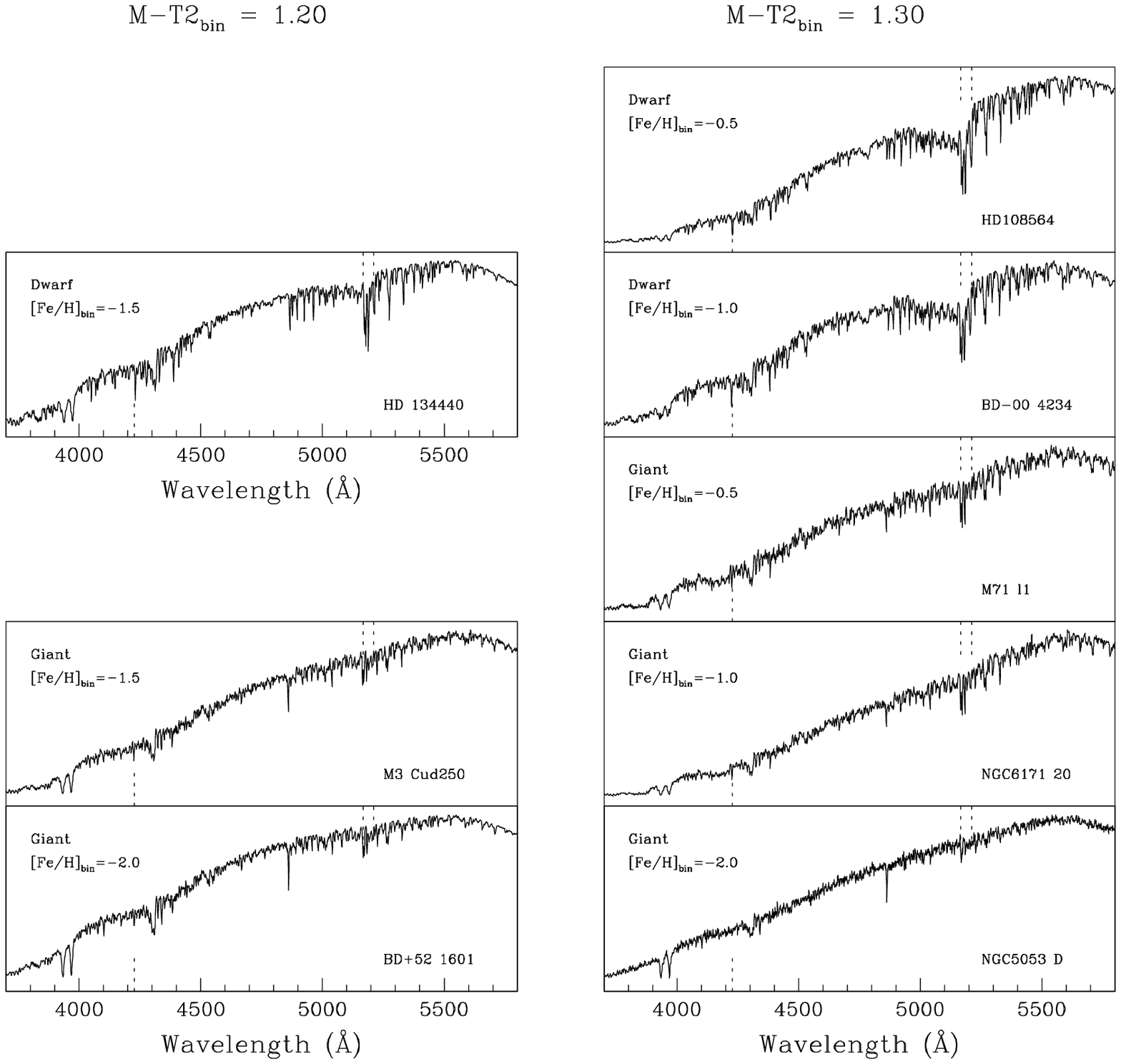,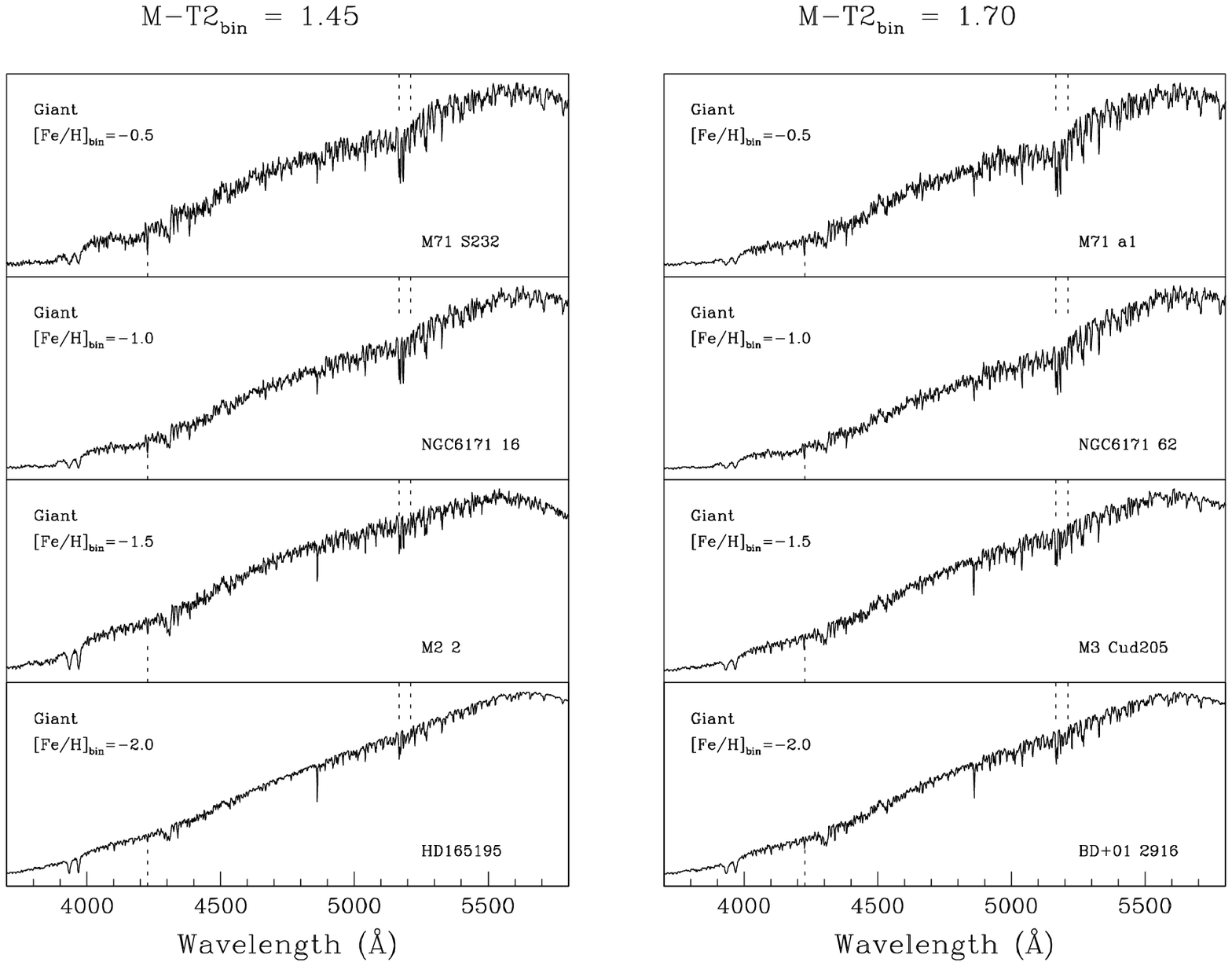]{Spectra of known late-type giants and dwarfs taken on the
KPNO 4m RC spectrograph in May 1999. Metallicity and \mi\ color of
star is given on each spectrum. Stars are ranked horizontally by \mi\
color, with the bluest stars to the left, and by both luminosity and
metallicity vertically. Dwarf spectra are shown at the top of each
vertical panel and giants at the bottom, with the most metal-rich in
each class at the top. The dwarfs in the bluer color range show a weak MgH
feature because they are hotter than the temperatures where the MgH
band is strongest. For stars with \mi\ greater than 1.10, it can be seen that even
metal-poor dwarfs have a strong MgH feature, and can easily be
distinguished from metal-poor giants.  MgH bandhead, the \mgb\ line at 5167\AA\  and Ca I 4227\AA\ line are marked.
\label{Mgb/H}}
\end{figure}

\citet{doug84} augmented the original Washington system with the DDO
``51'' filter, an intermediate-band filter centered on the \mgb~ and
MgH features near 5200 \AA~, to give luminosity sensitivity for late G
and K
giants. The \mg\ color gives a photometric method of measuring the
strength of these features.  Figures \ref{Mgb/H}(a)--(d) show
the strength of the MgH feature in spectra of dwarfs of solar and
lower metallicity, and of the \mgb\ feature for giants of different
metallicity. Most of these stars do not have direct measures of \mi\
color. We transformed to Washington colors from existing \by\ or \bv\
colors, using the method discussed in the Appendix.  For ease of
display we have sorted the stars into bins of metallicity and
color. Table \ref{lumrefs} gives sources of metallicity and color for
all the stars shown in Fig. \ref{Mgb/H}.

These spectra clearly illustrate the luminosity indicators discussed
above. A strong MgH band is seen for dwarfs with \mi\ = 1.2. The feature
becomes weaker as temperature increases, until it is hard to see in
the dwarfs with \mi\ = 1.0. Although we do not have any spectra for
dwarfs redder than \mi\ = 1.3, the MgH band continues to strengthen
with decreasing temperature. For all except the reddest giants (\mi$
>1.45$) there is no MgH feature visible, and the \mgb\ lines are
weak. For \mi of 1.45 and redder, there is a slight MgH feature
visible for the giants with [Fe/H] = --1.0 and above.

Our main method of detecting foreground dwarfs photometrically is via
these features (MgH + Mg b).  Other spectral features which are useful
for luminosity discrimination with a follow-up spectrum are the Ca I
4227\AA~ line and the CN bands.  The Ca I 4227\AA~ line is visible in
all the dwarf spectra, and can be seen to increase in strength as
temperature decreases. It is also visible in the spectra of the
metal-poor giants (especially above [Fe/H] = --1.5) but is much weaker
than in dwarfs of the same color.  Blue and UV CN bands are visible in
the metal-richer giants (for example M71 stars l-1, S232) and will be
discussed in more detail in the next section.

The 51 filter gives us the ability to measure the strength of the MgH
feature and reject the numerous foreground dwarfs. Precise photometry
is necessary, however: \cite{doug84} obtained \mg\ colors for a
sample of metal-rich and metal-poor giants, and 
predominantly metal-rich dwarfs.  He showed that giants and
metal-rich dwarfs differ by at least 0.10 mag in M-51 color for \bv$
>$ 0.85, and that the difference becomes more pronounced between
metal-poor giants (with weaker \mgb) and metal-rich dwarfs.

Figure \ref{m51mt2} shows \mg colors for known dwarf and giant
stars. \citet{doug84} plotted \mg\ versus $T_1-T_2$; it is clear from
Fig. \ref{m51mt2} that
the use of \mi\ as a color does not degrade the luminosity
sensitivity.  Our measurement errors for \mg\ are 0.02 to 0.04 mag.,
allowing us to discriminate easily between giants and all dwarfs
except the most metal-poor using our photometric survey.

Because we are searching for intrinsically very rare objects, we are
particularly vulnerable to photometric errors --- dwarfs with 3-sigma
errors in their colors are more common than our most distant halo
giants.  Since all but a few percent of the known halo stars have
[Fe/H]$<$--1.0, we can use metallicity as an additional criterion. We
use an additional Washington filter to identify halo giants: C. \cm\
is a metallicity indicator which was calibrated for giants by
\citet{doug91}.  We require that candidates lie in the [Fe/H] $<-1.0$
region of the Washington \cm\ vs. \mi\ diagram.

Figures \ref{m51mt2.4m} and \ref{cmmt2.4m} show how successful our
photometric classification has been. Photometric data from 22
high-latitude fields observed at the CTIO 4m with the BTC in April
1999 are plotted with spectroscopic confirmations shown as larger
symbols. These data will be described in more detail by
\citet{robbie}.  Note that we deliberately chose to observe candidates
near the giant/dwarf boundary to mark it carefully for future work.

\subsubsection{Extreme K subdwarfs}

Metal-poor halo dwarfs are of particular concern because their spectra
more closely resemble metal-poor giants. The Bahcall-Soniera model
\citep{bsmodel} predicts that 15 halo dwarfs will be found per square
degree down to V=20 in the color range that we search for giants (\bv
= 0.9 to 1.2). (Recall from Section 3.1 that we expect to see 4-5 halo
giants in the same magnitude interval.)

There are very few metal-poor K dwarfs with \mg\ photometry available, so we
cannot accurately measure the photometric separation between
metal-poor dwarfs and giants. However, \cite{pb94} have calculated
synthetic \mg\ colors for their grid of dwarf and giant models of
different metallicity. These models are based on the linelists of
\citet{bell94}. We show their dwarf sequences for [Fe/H] from 0.0 to
--3.0 in Fig. \ref{m51mt2}. It can be seen that the most metal-poor
dwarfs, with [Fe/H] $<$ --2.0, overlap the region where giants are
found in this diagram. Thus, the models suggest that the \mg\
photometric index will not be useful for weeding out the extreme K
subdwarfs in our fields --- we will need to examine their
spectra in more detail.

How common are these extremely metal-poor subdwarfs?  We can estimate
the number of halo dwarfs with [Fe/H] $<$ --2.0 using the halo
metallicity distribution of \citet{sean91}; 31\% of their sample has
[Fe/H] $<$ --2.0, which, in conjunction with the results of the
Bahcall/Soniera model, translates to 4 very metal-poor halo dwarfs per
deg$^2$ for $V<20$ at the NGP. Halo dwarfs only appear in significant
numbers for $V>18$, but the very distant halo giants also have
$V>18$. We need to consider these contaminants seriously.

We have approached the problem of discriminating spectroscopically
between halo giants and extreme K subdwarfs in two ways --- by
obtaining spectra of the few extremely metal-poor K dwarfs known in
this temperature range, and supplementing these with synthetic
spectra. It is practical to depend on follow-up spectroscopy to weed
these objects out because of their rarity.

Since some of the major differences between K dwarfs and giants are
due to molecular bands, we have used the NextGen model
atmosphere grid \citep{haus99a,haus99b} for the synthetic
spectra. These models were originally designed to model the
atmospheres of very low-mass stars and brown dwarfs, and include a
detailed molecular equation of state and a set of molecular opacity
sources which improve on those used in previous work. \citet{haus99a}
state that their models are more suitable for stars with $T_{eff} <
5000$K than previous models such as that of \citet{kurucz92}. 

The NextGen models are described in \citet{haus99a,haus99b}. Peter
Hauschildt kindly computed NextGen models with [Fe/H] = --2.0 and an
over-abundance of the alpha elements similar to that seen in halo
stars ([$\alpha$/Fe]=0.4) for us. They were calculated with log g =
1.0 and 4.5 to match giants and dwarfs with effective temperatures of
4700 and 4500K.  The model spectra are shown in Fig. \ref{lum_models},
smoothed to the same resolution as our 4m spectra. It can be seen that
there are marked differences between giants and dwarfs with
[Fe/H]=--2. Both dwarf models, especially the cooler one, show MgH
features, and both show strong lines of Ca I 4227, unlike the giants,
and much stronger \mgb\ lines. The dwarfs also show much stronger
lines in the region blueward of the Ca H and K lines, particularly in
the region near 3840 \AA~ where lines of FeI and MgI contribute.

In the next two figures we focus on two regions of the spectrum which
are particularly useful for luminosity discrimination, comparing
spectra of known metal-poor dwarfs and giants and supplementing with
synthetic spectra when no real spectra are available.

Fig \ref{ksubdblue} shows the region between 3700 and 4500\AA~. It can
be seen that the most metal-poor dwarf for which we have a spectrum,
HD 134440 with [Fe/H]=--1.5 and T$_{eff}$=4740K \citep{carbon} matches
the model spectrum with [Fe/H]=--2.0 and T$_{eff}$=4700K quite well,
giving confidence in the synthetic spectra. There is no marked
difference in G band strength between giants and dwarfs, but all the
dwarf spectra show a strong feature at Ca I 4227. The metal-poor
giants of moderate metal deficiency ([Fe/H] $\geq$ --1.6) show a
weaker Ca I line, while the very metal-poor giant shows almost none.

The blue and UV CN bands are also visible in the giants with [Fe/H]
$\geq$ --1.6, while the dwarf spectra look quite different in this
region. The feature just blueward of 3840\AA~ is very strong in the
dwarfs, and is significantly narrower than the UV CN band.
We note that because of CN anomalies in globular clusters, the CN
strength of the stars in Fig  \ref{ksubdblue} may not be typical of
field stars. While CN measurements of the three globular cluster
giants shown in the Figure are not available, field giant stars
do show these CN features, as can be seen in Figure 4(b) of
\citet{fm90}. This criterion can be used as a way of
confirming that a star is a giant, because no dwarf shows these CN
features. It cannot be used as a way of confirming that a
star is a dwarf because some giants may be CN weak.

Fig. \ref{ksubdmg} shows the region from 5000 to 5300\AA~. While the
large-scale shape of the MgH band is not easily visible when such a
small region of spectrum is displayed, it can be seen that at a given
\mi color, both the MgH bandhead and the Mg triplet are stronger in
the subdwarfs than in any of the giants.  Even the M71 giants, whose
metallicity is higher than we aim to identify in our halo sample, are
distinguishable from subdwarfs.
However, the effect is too small to depend on the \mg\ color to identify
these subdwarfs reliably. 

\vskip 0.7cm
 
In conclusion, our photometric survey will weed out the more common
dwarfs of the thin and thick disk, but is unable to identify the
very metal-poor K subdwarfs of the halo. These stars are rare enough
that it is practicable to use follow-up spectra with good S/N to reject
them.
We can measure our success rate for photometric
pre-selection using the percentages of spectroscopically
confirmed giants, foreground dwarfs and subdwarfs. 70\% of giant
candidates in the correct region of the \mg\ vs.\mi\ diagram were
giants, 20\% were subdwarfs and 10\% were dwarfs. 
Given our expected lack of success in discriminating subdwarfs from
giants photometrically, this confirms that our photometric selection
technique is very effective.

Figure \ref{mwgiants} illustrates the success of our selection with spectra
of metal-poor giants identified in our survey. Metallicity and
distance values obtained from the Washington \cm\ vs. \mi\
calibration are given for each star. Our sample already includes two
stars which are more distant than the LMC, one of which is shown in
Fig. \ref{mwgiants}, and we are well placed to
discover a large number of giants in the outer halo where only dwarf
satellites and globular clusters were previously known.

\subsubsection{Extragalactic Contaminants}

Possible extragalactic contaminants of our giant sample are QSOs and
unresolved galaxies.  
\cite{inr93} show
that there are $\sim$ 15 QSOs and unresolved galaxies \perdegsq\ down
to R=19 at the NGP. 

Most QSOs are separable photometrically because of their unusual
positions in the color-color diagram, although a few need to be weeded
out spectroscopically. However, color discrimination does not work for
unresolved normal galaxies, as their integrated colors are similar to
those of metal-poor giants \citep{doug95}.  The large pixels on the
Schmidt telescopes mean that we are more vulnerable to this problem
there.  With smaller pixels the galaxy contamination is much less
severe --- on our recent 4m run (reported in \cite{edo99}) where 
fourteen halo giants were identified from CTIO 4m/BTC data, no galaxies were
mistakenly found.

\subsection{Blue Horizontal-Branch Stars}

These stars will cover a distance range of 5 -- 50 kpc, and thus
represent another important tracer of the outer halo in our
survey. The number of BHB stars per \degsq\  depends on the
horizontal-branch morphology of the halo field, which is not well
determined for large distances from the Galaxy's center. If the halo
field follows the globular clusters in having a redder horizontal
branch morphology for large radii, then we would expect a few BHB
stars per \degsq\ in the magnitude range $V$=15--20.

We restrict ourselves to the portion of the horizontal branch which is
flat in V magnitude, between \bv=0.0 and 0.20. This converts to a
color range of 0.0 to 0.30 in \mi\ (see Fig. \ref{mt2bv}).
 
It can be seen from Figure \ref{cmd} that this is a sparsely-populated
portion of the color-magnitude diagram, and the only non-halo
contaminants of our sample in this color range are white dwarfs and
QSOs.  There are 3 white dwarfs \perdegsq\ and 5 QSOs \perdegsq\ down
to V=20 in this color range, \citet{tom86,sand69,inr93}. Both of these
types of
objects are easy to discriminate with even a low-dispersion, low
signal-to-noise spectrum.

 We also need to be able to discriminate between halo blue stragglers
(with main sequence gravities) and blue horizontal branch stars.  Our
followup spectroscopy allows us this via Balmer line profiles
\citep{pier83,slc,nh91,tdk94}.  In cases where we are able to obtain
spectra which reach below the Balmer jump, its size can also be used
as a discriminator. \citet{tdk94} used spectra of 3.7\AA~ resolution
(very similar to our 4m spectral resolution) to make this measurement.

Figure \ref{bhbspectra} shows spectra of known BHB standards and two
BHB stars from our sample.

\subsection{Halo turnoff stars}

Halo turnoff stars are the most numerous but least luminous tracers we
shall use, and sample distances from 2 -- 16 kpc from the Sun
(assuming an absolute magnitude $M_V$ = 4.5 and a limiting magnitude of
V=20.5).

We used the halo turnoff luminosity function calculated in Section 4.1
and the preferred halo model from Section 4 (power law exponent of
--3.0 and an axial ratio of 0.6) 
to calculate the numbers of turnoff stars per square degree
we would expect to see in our fields. Under these assumptions,
there should be 150 halo
turnoff stars per square degree down to V=20.5 at the NGP, and 130 per
square degree in the anticenter at $b$=45.

These stars are relatively easy to identify using accurate photometry
because they are bluer than almost all stars at high galactic
latitude. For the most metal-poor globular clusters such as M92, the
turnoff is at \bv=0.38, significantly bluer than the turnoff color of
the thick disk (\bv=0.5, \citet{carn89}).  The thin disk has a
sufficiently small scale height that very few young thin disk stars
are found in our magnitude range at high galactic
latitude \citep{bsmodel}.  We choose stars with \bv\ color of 0.38 to
0.49 as halo turnoff star candidates.  

The transformation to \mi\ color is complicated by the fact that we
have no observations of metal-poor turnoff stars in this color. We
approach the problem in two ways. First, only the most metal-poor
stars will have a turnoff color as blue as \bv\ = 0.38, so we could
use the synthetic colors of \citet{pb94} for their models with [Fe/H]
= --2.0 to derive the transformation. This predicts a turnoff color of
\mi=0.6. Second, we can derive the turnoff color using \vi\ and Stromgren
photometry of metal-poor globular clusters from the literature. Table
\ref{turnoffcolors} summarizes the turnoff colors in \vi\ and \by\ of
metal-poor globular clusters NGC 6397, NGC 7099 and M92, and the
metal-rich cluster 47 Tucanae.
The average \vi\ turnoff color for these three metal-poor clusters is
0.51, which transforms via Fig. \ref{mt2vi} to \mi=0.64.
Thus the  estimates from theory and observation are in
reasonable agreement, and we have chosen to use the observational estimate
here. We also use the \vi\ photometry of \citet{kaluz} for 47 Tuc to
constrain the turnoff color of the thick disk to be \mi=0.81.  In
summary, we take the color range from \mi\ = 0.64 to
0.80 for our halo turnoff stars.
As a consistency check on our photometry, we also require that  \cm\
is in the range 0.2 to 0.5.

Thick disk stars scattering into the color range should be a minor
problem because of our small photometric errors (our \mi\ errors are
less than 0.02 mag. at V=19) and the fact that we work faint enough to
be away from the regions of the Galaxy where the thick disk dominates.
Using the models of turnoff stars in the thick disk and halo of
section 3.1, we can show that only for the brightest part of our
magnitude range (V=16--17, corresponding to distances of order 2.5
kpc, saturated on 4m/BTC data but not on Schmidt data) are numbers of
thick disk stars per square degree greater than halo star numbers in
our fields, and then only by a factor of $\sim$2, which is not large
enough to make significant numbers of thick disk stars ``leak'' into
the halo color range via photometric errors. For stars with V=17--18,
numbers of thick disk and halo turnoff stars are approximately equal,
and for fainter magnitudes, halo turnoff stars outnumber thick disk
stars.

The only other contaminants of our sample in this color range are
small numbers of white dwarfs, QSOs and RR Lyraes. The QSOs and white
dwarfs are obvious from spectra, and the RR Lyraes are sufficiently
rare that few will be detected and some of these will be removed
due to their variability on either photometric or
spectroscopic observations.

Figures \ref{cmd} and \ref{turnoffs} illustrate the selection
technique, showing color-magnitude diagrams (\mi\ vs M) for a single
field and for a composite of many fields. The position of the turnoff for both halo
and thick disk is marked, and our candidate halo turnoff stars are
shown. With very few exceptions, all stars in this region of the CMD
show spectra typical of halo stars, i.e. metal-poor.  Figure
\ref{turnoffstars} shows WIYN/Hydra spectra of several of these stars.
It is clear
that the halo turnoff candidates are indeed metal-poor, confirming the
accuracy of our photometry.

\subsection{Blue Metal-Poor Stars}

\citet{pbs94} identified an important group of halo turnoff stars with
color bluer than \bv\ = 0.38, and suggested that these stars were
metal-poor stars with unusually young ages, which had originated in
dwarf spheroidal satellites which were subsequently accreted into the
Galaxy's halo. Another possibility is that these stars are the result
of the evolution of multiple star systems of the halo. Preston has
obtained detailed follow-up spectroscopy of a number of these stars to
test this possibility. \citet{pbs94} estimated numbers in the solar
neighborhood of 350--450 per kpc$^3$ (cf. the density from Table
\ref{halolf} of 2695 halo turnoff stars per kpc$^3$ with $M_V$=4.5).
\citet{unavane} found from photographic starcount data that $\sim10$\%
of halo stars were bluer than \bv\ = 0.4, which is in rough agreement
with the \citet{pbs94} value. These BMP stars are particularly important
for our survey if they have younger ages, and we select them by
\bv\ color between 0.15 and 0.35 (\mi\ = 0.2 to 0.6).

Using the \citet{pbs94} local normalization, we expect to find 10--20 BMP stars
\perdegsq\ to V=19. Our Fig \ref{cmd} shows 23 
such stars in the color range \mi\ = 0.2 to 0.6 for an area of 2.75
\degsq, a little lower but not significantly different from the value
of \citet{pbs94}.

\section{MAPPING THE HALO -- ``INTELLIGENT'' STAR COUNTS}

Accurate CCD photometry of large areas in many fields makes an new
method of investigation of the halo possible. As discussed in Section
3.3, the halo turnoff color is \bv = 0.38, and the thick disk turnoff
is \bv $\sim$ 0.5, so stars with colors between these two numbers are
almost certainly halo stars close to the turnoff. Photographic colors
have such large errors (0.05 to 0.10 magnitudes) that it is not
possible to isolate halo turnoff stars from photographic data without
simultaneously modelling the contribution of the thick disk.  In
contrast, our photometry is of such high and uniform quality that it
is possible to separate halo turnoff stars cleanly from thick disk
turnoff stars using the \mi color. Fig. \ref{turnoffs} illustrates
this in one of our lowest latitude fields, where both thick disk and
halo turnoffs can be seen.

We now make a preliminary analysis of the BTC data to
illustrate the power of our survey technique. Our major aim here is to
check whether our data agree with other models for the halo.

The dataset obtained with the Big Throughput Camera on the CTIO 4m in
April 1999 \citep{robbie} is particularly useful for mapping the halo
because of the uniform quality of the data, and the fact that
conditions were photometric throughout the run. Fig. \ref{cartoon}
shows in cartoon form the location of the fields where data were
obtained. There are 46 fields with latitudes ranging from +25 to +73,
and longitudes from $l$=17 through the galactic center to $l$=218
(less than 40 degrees from the anticenter).

We have checked for errors in reductions or reddening estimation by
carefully examining the color-magnitude diagrams of all fields.
Figure \ref{turnoffs} is typical: the position of
the halo turnoff is clear, and agrees well with the calculated value
of $(M-T2)_0$=0.64 in all but two of the 46 fields.  In these two
fields it appears that the Schlegel reddening values that we used
should be adjusted by a few hundredths in order to bring the turnoff
position to this color.  Because the stars are uniformly distributed
across the color range from $(M-T2)_0$=0.64 to 0.8, photometric errors
or reddening estimates of a few hundredths of a magnitude will in no
case be strong enough to make the turnoff star numbers vary
significantly. 

Models of the shape of the halo have in many cases been derived from
easily detectable tracers such as globular clusters \citep{zinn} or
horizontal-branch stars \citep{kin65,abi,hart87,psb91,tdk94}. Studies
using star counts (eg Bahcall and Soniera 1986) are handicapped by
their inability to separate halo and thick disk accurately with
photographic photometry, as discussed above.  Models from both
globular clusters and field stars both find that the halo is centrally
concentrated, with power-law exponent varying from --3.0 to --3.5 or
even steeper. Axial ratios vary from 0.5 to 1.0, with suggestions from
several groups that the axial ratio might change with galactocentric
radius in the sense that the outer halo is spherical and the inner
halo flattened.

However, neither globular clusters nor horizontal-branch stars are
ideal for measuring the density distribution of the halo. First, it is
not clear that the halo field stars were formed under the same
conditions as the globular clusters. Second, possible age and
metallicity gradients in the halo \citep{szinn, zinn93, psb91, tdk94}
are reflected in horizontal-branch morphology. This can cause a
different power-law exponent to be derived for RR Lyraes and BHB
stars, as found by \citet{psb91}. 
Thus a check of the earlier results with a different
tracer is valuable. Although our turnoff star sample will
also be sensitive to age and metallicity variations, we are  able
to check for such effects by examining the color-magnitude diagrams of
discrepant fields directly.

\subsection{Local Halo Density}

The estimation of the halo density in the solar neighborhood is even
more challenging than the measurement of its density in more distant
fields.  It has in most cases been based on the local density of stars
selected by their high proper motion (eg Bahcall and Casertano 1986),
with kinematic corrections made for the amount of the halo that would
be missed using this selection technique. There have also been a small
number of direct measurements of local density of some tracer such as
RR Lyraes or red giants (eg Preston et al. 1991, Morrison 1993) which
agree within a factor of two with the proper-motion data, but
different selection effects such as metallicity operate for these
samples. Attempts to extrapolate the results of pencil-beam
surveys inward to the solar neighborhood are not always successful ---
\citet{psb91} and \citet{wett96} note the
disagreement of a factor of two between the counts of nearby RR Lyrae
variables and the extrapolation of more distant RR Lyrae counts
inward.

Since we have few low-latitude fields and our BTC data saturate for
magnitudes much brighter than V=17 (corresponding to a distance of 3
kpc for turnoff stars) we have no direct constraints on
the local density from our data. Our preferred value of the solar
neighborhood halo density will depend on the axial ratio adopted.

We decided to re-examine the measurement of the local halo density
from the proper-motion samples, as there have been significant
advances in the information available on these stars since the work of
\cite{bc86}. We have used the extended sample of \citet{carn94}, which
has been updated recently to have a distance scale consistent with the
Hipparcos parallax measurements for subdwarfs, to make an estimate of
the local density. Bruce Carney kindly made this sample available to
us in advance of publication.  Also, in order to make comparisons with
our BTC data easier, we isolated stars in the turnoff star color bin
we used (\bv\ between 0.38 and 0.5, corresponding to \mi\ between 0.64
and 0.8), and then derived a local luminosity function for these stars
alone.  Since this is a kinematically selected sample, we need to make
corrections for the halo stars missed because of the proper motion
selection. To minimise contamination of the sample by thick disk
stars, which contribute strongly to the derived local density because
the lowest velocity stars are given the highest weights (see Bahcall
and Casertano 1986) we only used stars with tangential velocity
greater than 220 km/s, and rejected all stars with [Fe/H] $>$ --1.0.
This may reject a few genuine halo stars, but will cause an error of
only $\sim$10\% in the derived density \citep{carn89}. Erroneously
including thick disk stars would have a much larger effect.

The \citet{carn94} sample is drawn from the Lowell Proper Motion
Catalog which covers the entire Northern Hemisphere and has a proper
motion lower limit of 0.26 $''$ yr$^{-1}$. In the color range of
concern here, the catalog's magnitude limit is sufficiently faint that
all we need to do is correct for proper motion selection effects,
which we do by weighting by $V_{tan}^{-3}$, following \citet{bc86}.
Table \ref{halolf} gives our results. Carney and collaborators have a
more complete kinematical analysis in progress, so we have chosen
simply to use the simulations of \citet{bc86} to correct for the
tangential velocity cut at 220 km/s. \citet{hlm93} notes that the
sample used by Bahcall and Casertano to derive halo kinematics was
probably contaminated by thick disk stars, and calculates that the
``discovery fraction'' for a sample with this $V_{tan}$ cutoff should
be close to 0.5, not 0.33 as they found. We have used this higher
value in our calculations of the turnoff star luminosity
function.

\subsection{Halo concentration and axial ratio}

While our maps of the halo based on accurate color-magnitude diagrams
are less sensitive to possible variations in age and metallicity than
the estimates from horizontal-branch stars in particular, they are
less easy to interpret because of the difficulty of obtaining accurate
distances to turnoff stars. Turnoff stars in our chosen color range
can have absolute magnitudes which vary from 3.5 to 5.5, although the
luminosity function of Table \ref{halolf} shows that many will have
absolute magnitude near 4.5. We have chosen to divide our data into
three magnitude ranges (V=17--18.5, 18.5--19.5 and 19.5--20.5) and to
calculate model predictions for these magnitude ranges to compare
directly with the star counts there.  We plan to do a more
statistically sophisticated analysis in a future paper --- the values
we derive here will be roughly correct for the distance ranges probed
by the data, but not may not be optimal.

Fig. \ref{axrat} shows the sensitivity of our data to the axial ratio
of the halo. It shows the ratio of number of stars in a given
magnitude bin to the model predictions. Both models have a power-law
exponent of --3.0, and one has an axial ratio b/a=1.0 (spherical)
while the other has a moderate flattening (b/a=0.6).  We have plotted
these ratios vs a rough indicator of the z height travelled by each
line of sight, calculated by assuming that all stars in the bin have
absolute magnitude $M_V$=4.5.  It can be seen in Fig. \ref{axrat} that
the spherical halo provides a significantly worse fit. There is a
trend with z: points with high z have fewer stars than the model
predicts, and points with low z have more. The b/a=0.6 model residuals
show little trend with z, and the small number of very discrepant
points remaining are found with both small and large z.

Fig. \ref{expon} shows the same data/model ratios against the range in
$R_{gc}$, for models with b/a=0.6 and power-law exponent --3.0 and
--3.5. A trend with $R_{gc}$ is visible in the panel with power-law
exponent of --3.5. The model predicts too few stars with large
$R_{gc}$ and too many with small $R_{gc}$. Note also that in the panel
showing residuals from the model with exponent --3.0, the remaining
residuals show no clear trend with $R_{gc}$.  We adopt a model with
b/a=0.6 and power-law exponent --3.0 for the rest of this
analysis. Only a 10\% correction to the local halo density of Table \ref{halolf} is needed
with these parameters.

It is possible to examine the residuals to the model fits in more
detail by plotting the position in the Galaxy traversed by each field,
and highlightling the fields with large residuals. 
Figures \ref{residz} and \ref{residr}  show a number of vertical and
horizontal  slices through the
Galaxy, for different ranges of y and z. Fields with residuals more
than 2.5$\sigma$ from the model fit are highlighted.  It can be seen that 
four of the fields with large negative residuals are towards the galactic
center, and two, with large positive residuals, 
are in our lowest latitude field, which is close
to the anticenter. Interestingly, the discrepant fields cover a large
range in both r and z, so a simple adjustment to the power-law
exponent or flattening will not improve things.
 
The variable axial ratio models of \citet{hart87} and \citet{psb91}
will not solve the problem completely. The model of \citet{psb91} has
an axial ratio changing linearly from b/a=0.5 at the galactic center
to b/a=1.0 for ellipsoids with semi-major axis of 20 kpc.  While the
results in fields close to the minor axis which currently have large
negative residuals may be improved by the adoption of a model where
the axial ratio is closer to 1 at this point, the anticenter fields
will become more discrepant since the axial ratio must become close to
1 at this distance from the center too.  However, it is possible that
the thick disk might be partially responsible for the excess of stars
in this color range, if its scale height increases in the outer Galaxy
and/or it has a strong abundance gradient.

We plan to investigate these issues further when we obtain more
fields with low values of $R_{gc}$ and z.

\section{CONCLUSIONS}

We describe a survey  designed to find  field stars from
the galactic halo in large enough numbers to provide a strong test of
the question ``Was the galactic halo accreted from satellite
galaxies?''

The survey will cover 100 \degsq\ at high galactic latitude. It
uses an efficient pre-selection technique based on the Washington
photometric system to identify halo red giants, blue horizontal branch
stars, blue metal-poor main sequence stars and turnoff
stars. Follow-up spectroscopy (with multi-object spectrographs for the
more numerous turnoff and BMP stars) tests for kinematic signatures of
accretion.  Our sample of halo stars will be unprecedentedly large,
and cover galactocentric distances from the solar circle to more than
100 kpc. Because the photometric selection has few and easily
quantifiable selection effects, our sample will also enable thorough
studies of the density distribution of the galactic halo, increasing
the numbers of distant halo objects known by an order of
magnitude. This will allow much more accurate measurement of the mass
of the Galaxy than previously possible.

We discuss the particular problems caused for identification of the
very distant halo giants by foreground K subdwarfs with very low
metallicity ([Fe/H]$<$--2.0). These stars are roughly as numerous as
the genuine halo giants for $V>18$, and are not distinguishable from
giants via the Washington \mg filter that we use for photometric luminosity
discrimination. However, with low-dispersion spectra of S/N 15 or
more, a combination of several spectral features such as the Ca I line
at 4227 \AA~ suffices to distinguish these stars reliably from giants.
Studies which use giants in the outer halo need to take particular
care to eliminate these extreme K subdwarfs.

We use one of our large photometric datasets, obtained in one run
using the BTC on the CTIO 4m, to constrain the spatial distribution of
the Galaxy's halo over ranges of galactocentric distances from
approximately 5--20 kpc and z heights from 2 to 15 kpc by using halo
turnoff star numbers. We find that a power-law with exponent --3.0 and
a moderate flattening of b/a=0.6 gives a good fit to most of the data.
However, there are a number of fields that show departures
from this model, suggesting that we will need a more complex model in
future as our coverage of fields in the halo increases.

\vskip 1cm \acknowledgements It is a pleasure to thank Peter
Hauschildt for kindly running the alpha element-enriched NextGen
models for us and making the NextGen models so readily available on
the Web.  We would also like to thank Doug Geisler for the careful
work he has devoted to making the Washington system so useful, and for
his help with our many questions about the system, and Bruce Carney
for kindly making his database available in advance of publication.
We have benefited from helpful discussions with Mike Bessell,
Bruce Carney, Conard Dahn and Peter Dawson. We would also like to
thank an anonymous referee for comments that improved this paper.

This work was supported by NSF grants AST 96-19490 to HLM, AST
95-28367, AST96-19632 and AST98-20608 to MM, and AST 96-19524 to
EWO. We used the SIMBAD database, maintained by CDS, Strasbourg,
extensively.

\newpage

\appendix{APPENDIX: CONVERSIONS BETWEEN WASHINGTON AND OTHER SYSTEMS}

The Washington broadband photometric system \citep{can76,hc79,doug96} includes the \ttwo\ filter (which is very
similar to the Cousins I filter), the M filter (which is 1050\AA~~
wide and has a central wavelength of 5100\AA, slightly blueward of the V filter)
and the C filter (which is 1100 \AA~\ wide, centered near the Ca K
line at 3934 \AA~). \citet{doug84} has added the DDO ``51'' filter to
make giant/dwarf discrimination possible for G and K stars. (The
system also includes the T1 filter, similar to the Cousins R filter,
which we do not need to use).

The original temperature indicator for the Washington system was the
$T_1-T_2$ ($R-I$) color. However, an alternative is the \mi\ color,
which transforms well to \vi\, as can be seen in Figure
\ref{mt2vi}. For \vi\ between 0.5 and 1.5, the simple linear relation
\mi\ = 1.264 (\vi) works well. The standard deviation of the residuals
from this line is 0.25 magnitudes.

We have included both dwarfs and giants in this diagram, and a
significant number of metal-poor globular cluster giants.
We used:
\begin{itemize}
\item \citet{arlo} standard stars which are also Washington standards
\citep{hc79, doug84, doug86, doug90, doug91,doug92,doug96}. Only stars
with E(\bv) less than 0.11 mag, from the reddening maps of
\citet{schlegel} are shown. We corrected the colors using the
relations between E(\bv) and the extinction in the Washington
passbands using the ratios from \citet{hc79}.
\item we used photometry of
stars from three globular clusters: 47 Tucanae, NGC 1851 and NGC 6752.
The \vi\ photometry is from \citet{garytaft}, and the
Washington photometry from \citet{doug86,doug97}.
\item Accurate photometry of a sample of very nearby dwarfs was
obtained by \citet{gpiche92}. They also measured \bv\ colors for these
stars. Although the authors did not measure \vi\ colors, \citet{msb79}
has shown that the Stromgren \by\ color transforms very well to \vi\,
and we have used his transformation, and the very accurate \by\
measurements of \citet{egg98}, to derive \vi\ colors for these
stars. We have used the six stars in their sample with $B-V \leq
0.51$ to supplement the Landolt standards. Because these stars are so
nearby (all have large and accurate parallaxes) reddening corrections
are not needed.
\end{itemize}

It is clear that there is no metallicity
dependence in the transformation from \mi\ to \vi.

It is also possible to transform the \mi\ color to \bv\, but this is
less straightforward because there are different loci for late-type
dwarfs and giants in this diagram. However, for bluer stars such as
halo turnoff stars, the relation is single-valued, as can be seen in
Figure \ref{mt2bv}.  Since there are few metal-poor dwarfs with
Washington photometry, we have also plotted the models of \citet{pb94}
which are based on synthetic spectra, for dwarfs with solar abundance
and [Fe/H] = --2.0. \citet{pb94} note that their models show good
agreement with existing data for temperatures higher than 4500 K
(\bv$\simeq$ 1.0).

Washington \cm\ is less useful as a temperature indicator because it
has both metallicity sensitivity for late-type stars (from line
blanketing around 4000 \AA~) and some gravity sensitivity for earlier
types (from the Balmer jump) We give the relation between \cm\
and \bv\ in Figure  \ref{cmbv} for
completeness.

It is also useful to derive transformations between the Stromgren \by\
color and \mi. This is particularly useful for K star temperatures,
because many of our spectroscopic luminosity standards have \by\
observations but not \mi. For main sequence stars, we used the
compilations of \citet{egg98,gpiche92} of photometry of stars from the
Yale Parallax Catalog \citep{yale}, whose stars have the advantage that
they are so close to the Sun that reddening corrections are unlikely
to be needed. For metal-poor giants we used stars from \citet{bond80}
with Washington photometry in \citet{doug86}. E(\bv) values were taken
from \citet{bond80}, and were smaller than 0.04 in all cases.

Figure \ref{bymt2} shows the relation between these two colors. It can
be seen that there are different sequences for red dwarfs and giants,
but that the sequences are fairly tight in both cases.

\newpage

\newpage
\begin{figure}
\includegraphics[scale=0.8]{Morrison.f3b.eps}
\end{figure}
\newpage
\begin{figure}
\includegraphics[scale=0.8]{Morrison.f3c.eps}
\end{figure}
\newpage
\begin{figure}
\includegraphics[scale=0.8]{Morrison.f3d.eps}
\end{figure}
\newpage

\clearpage

\begin{figure}
\plotone{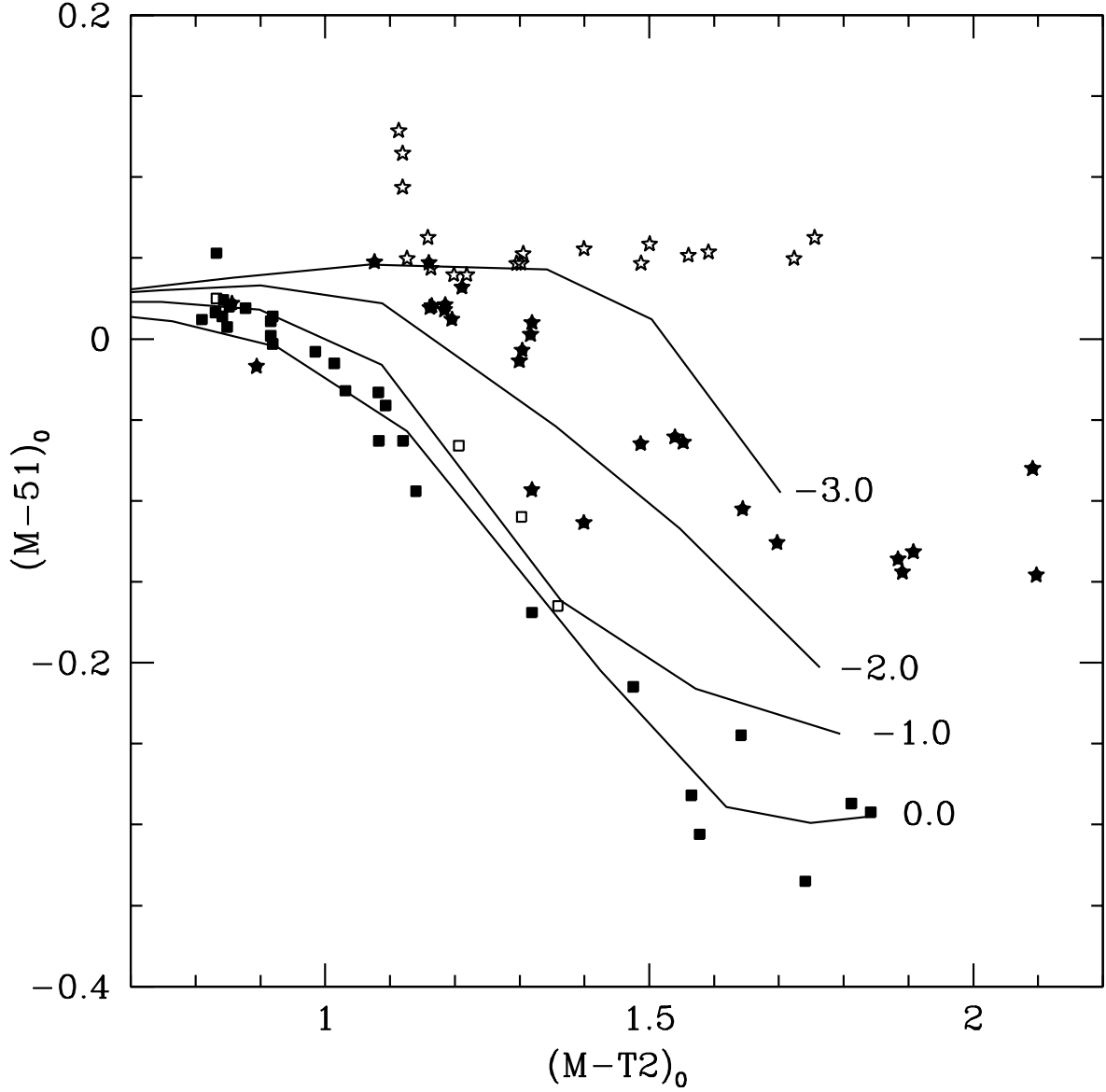}
\figcaption[Morrison.f4.eps]{\mg\ color for a selection of known late-type dwarfs and
giants plotted versus \mi\ color. Dwarfs with roughly solar
metallicity are plotted as filled squares, those with [Fe/H] between
--0.5 and --1.5 as open squares, solar abundance giants are plotted as
closed stars, and metal-weak giants from the globular clusters NGC
6541 ([Fe/H] = --1.83, Zinn 1985) and NGC 6397 ([Fe/H] = --1.91, Zinn
1985) are plotted as open stars. Loci traced by the dwarf models of
\protect\citet{pb94} for [Fe/H] from 0.0 to --3.0 are also shown.
\label{m51mt2}}
\end{figure}

\begin{figure}
\plotone{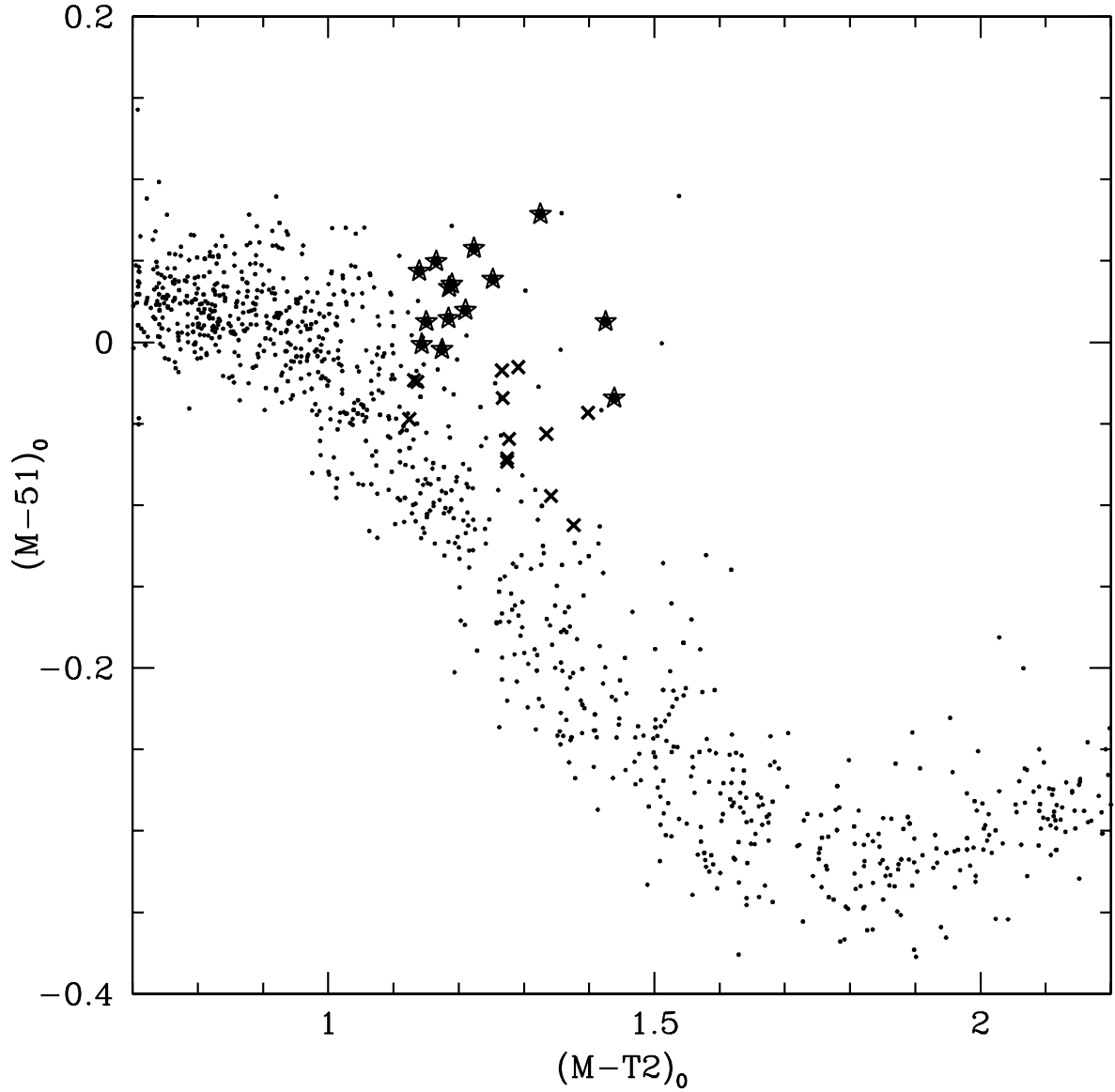}
\figcaption[Morrison.f5.eps]{Spectroscopically confirmed halo giant
stars in the \mg\ versus \mi\ diagram. Small circles are stars of all
colors identified on several BTC fields, plotted to show the region
where foreground dwarfs are found, crosses are halo giant candidates
that were found to be dwarfs when spectra were obtained, and stars are
spectroscopically confirmed halo giants. We deliberately observed
stars with low \mg\ values to delineate the dwarf region as accurately
as possible. 
\label{m51mt2.4m}}
\end{figure}

\begin{figure}
\epsscale{0.7}
\plotone{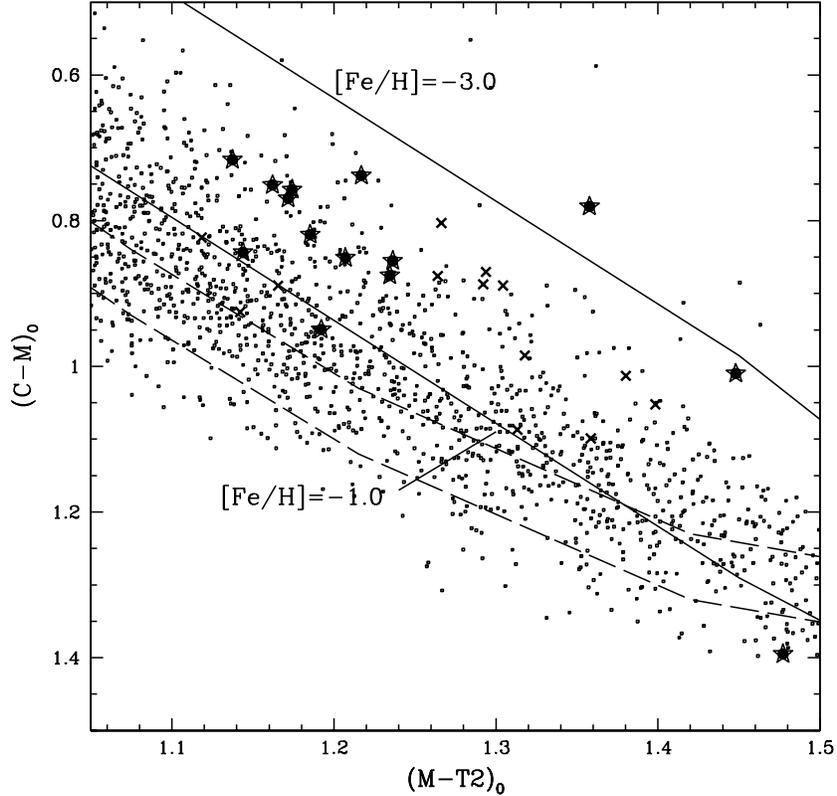}
\figcaption[Morrison.f6.eps]{Spectroscopically confirmed halo giant
stars in the \cm\ versus \mi\ diagram. Small circles are stars of all
colors identified on 8 BTC fields, plotted to show the region where
most foreground dwarfs are found, crosses are halo giant candidates
that were found to be dwarfs when spectra were obtained, and stars are
spectroscopically confirmed halo giants. Solid lines show the
calibration lines for [Fe/H]=--1.0 and --3.0 from \protect\citet{doug91} and
dashed lines show the calibration lines of \protect\citet{gpiche92} for
[Fe/H]=0.0 and --0.5. Since the metal-poor giants have a smaller
separation from the dwarf sequences for cool stars, identification
becomes less efficient there. It can be seen that the model prediction
of roughly equal numbers of thick and thin disk dwarfs in our fields
is borne out by our data: few stars are found below the [Fe/H]=0 line
for dwarfs, and there are significant numbers of stars near the [Fe/H]
=--0.5 line. [Fe/H]=--0.5 should represent the mean abundance for
thick disk dwarfs.
\label{cmmt2.4m}}
\end{figure}
\clearpage

\begin{figure}
\plotone{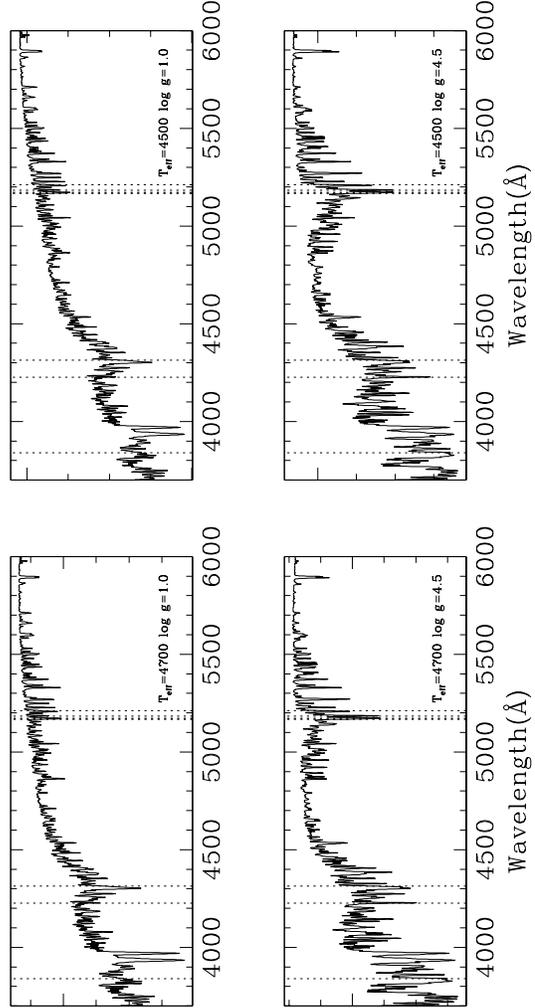}
\figcaption[Morrison.f7.eps]{Model spectra for stars with [Fe/H] = --2.0,
[$\alpha$/Fe]=+0.4, log g = 1.0 and 4.5 and T$_{eff}$=4700 and 4500K.
Positions of the MgH bandhead at 5211\AA~, the Mg triplet, the G
band, the CaI 4227\AA~ line and the group of FeI and MgI lines
near 3840\AA~ are marked.
\label{lum_models}}
\end{figure}

\begin{figure}
\plotone{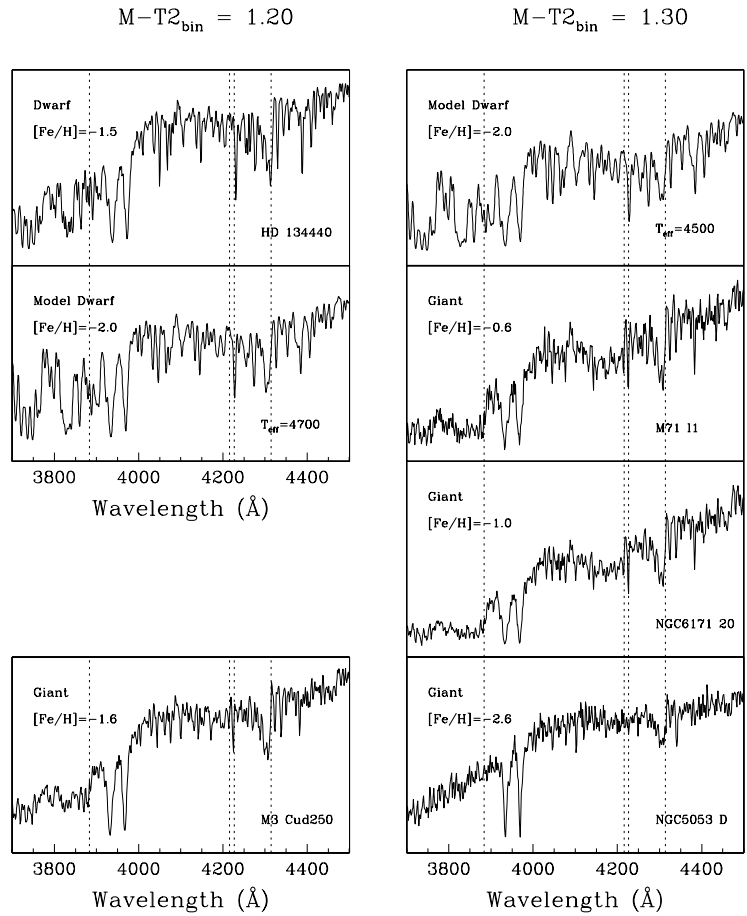}
\figcaption[Morrison.f8.eps]{Spectra of metal-poor subdwarfs and
giants of various metallicities, sorted into \mi bins, in the spectral
region from 3700 to 4500\AA~. Wavelengths of the G band, Ca I 4227
line, blue CN bandhead and the UV CN bandhead are marked. HD 134440
has a radial velocity of 308 km/s, which explains why its spectral
lines are offset from their rest wavelength.
\label{ksubdblue}}
\end{figure}

\begin{figure}
\plotone{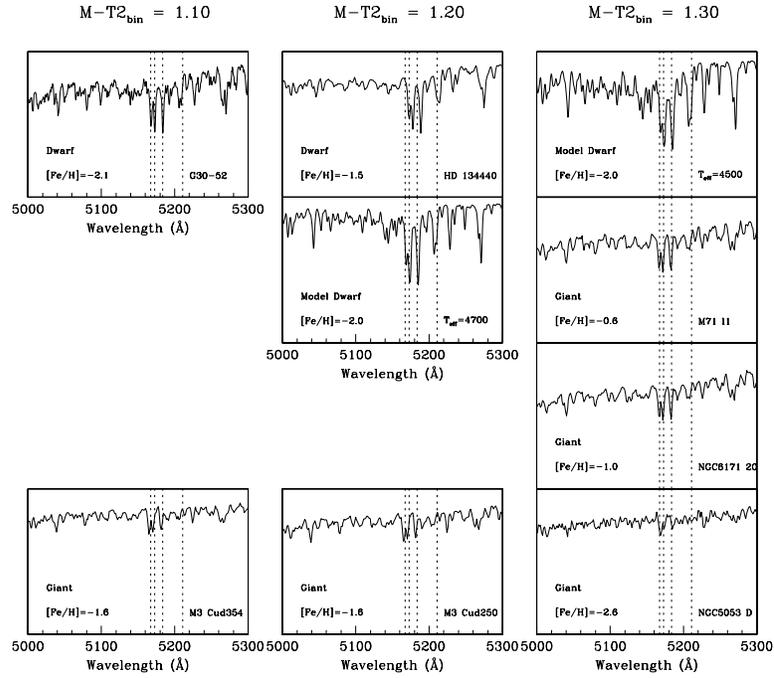}
\figcaption[Morrison.f9.eps]{Spectra of metal-poor subdwarfs and
giants of various metallicities, sorted into \mi bins, in the spectral
region from 5000 to 5300\AA~. Wavelengths of the MgH bandhead at
5211\AA~ and the Mg triplet are marked.
\label{ksubdmg}}
\end{figure}
\clearpage
\begin{figure}
\epsscale{0.75}
\plotone{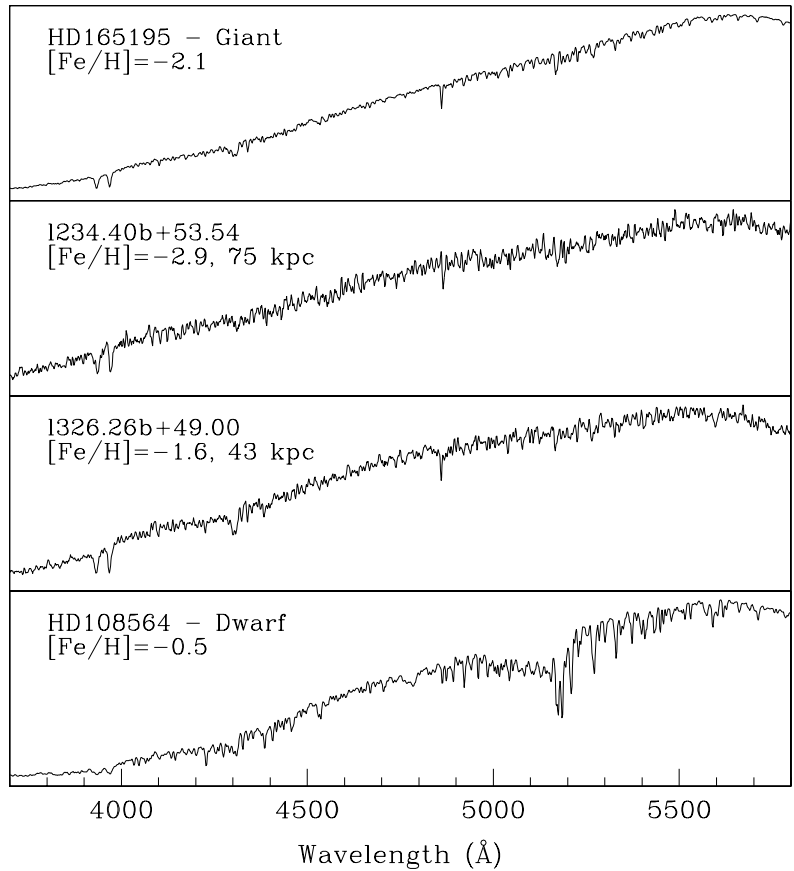}
\figcaption[Morrison.f10.eps]{Spectra of two metal-poor red giants
identified in our survey, compared with a known giant and dwarf. These
spectra were obtained with the KPNO 4m RC spectrograph in May 1999 and
will be described more fully in \protect\citet{edo99}. The star name gives its
galactic longitude and latitude. The metallicity and distance of the
star is given on each spectrum.
\label{mwgiants}}
\end{figure}

\begin{figure}
\epsscale{0.75}
\plotone{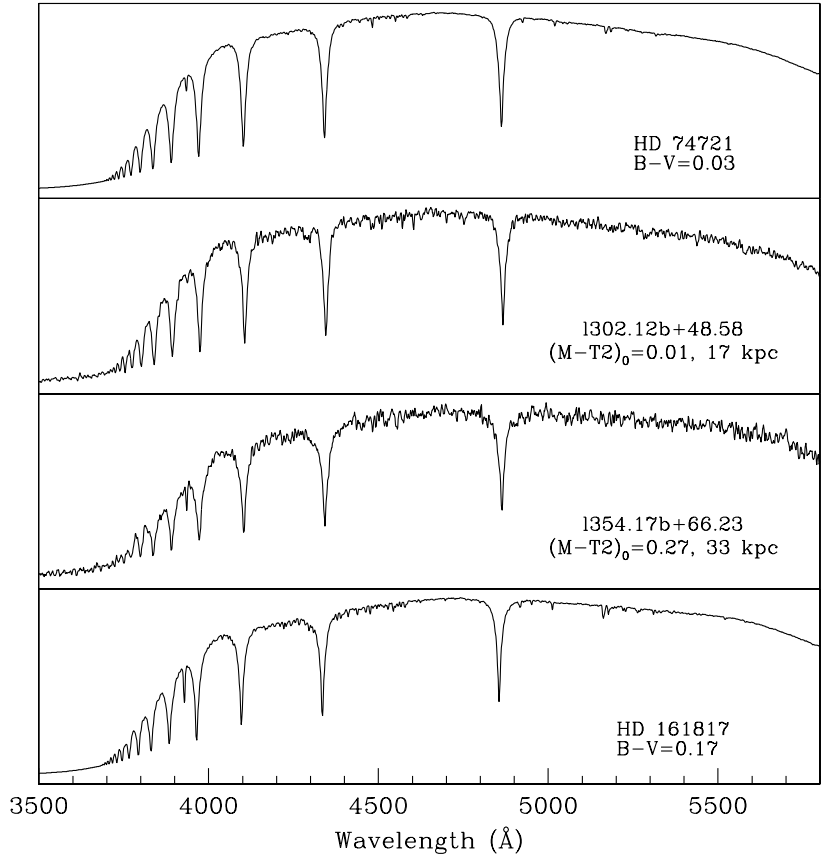}
\figcaption[Morrison.f11.eps]{Spectra of BHB stars identified in our
halo fields, compared with spectra of known field BHB stars. Stars are
plotted in order of decreasing temperature. These data will be
described more fully in \protect\citet{edo99}.
\label{bhbspectra}}
\end{figure}

\begin{figure}
\epsscale{0.75}
\plotone{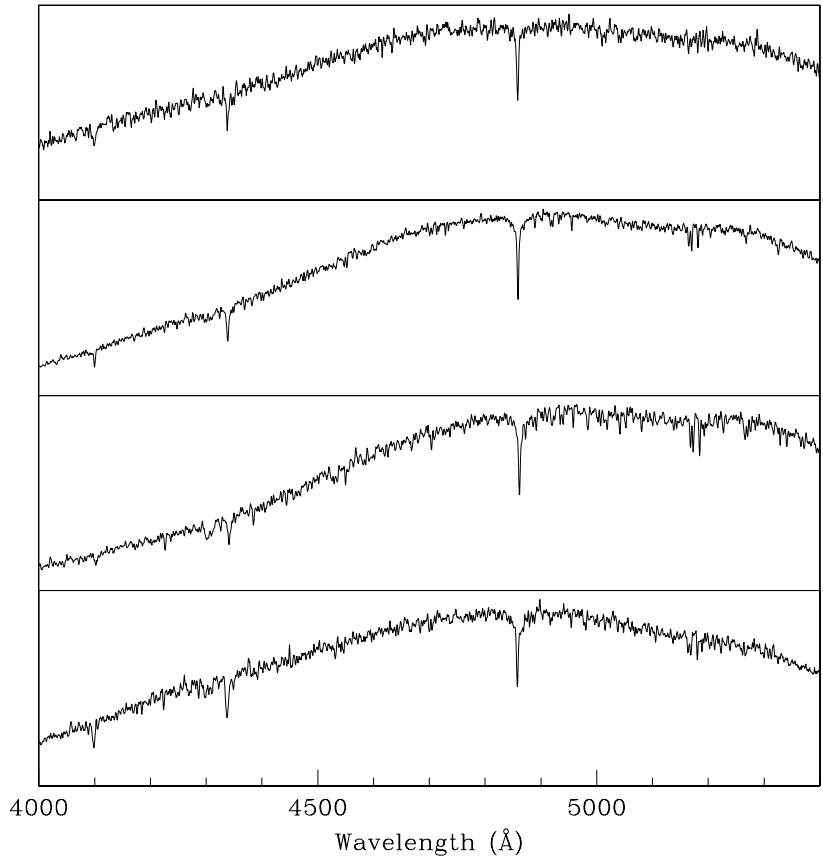}
\figcaption[Morrison.f12.eps]{WIYN/Hydra spectra of main sequence
turnoff stars identified in our halo fields.
\label{turnoffstars}}
\end{figure}

\begin{figure}
\plotone{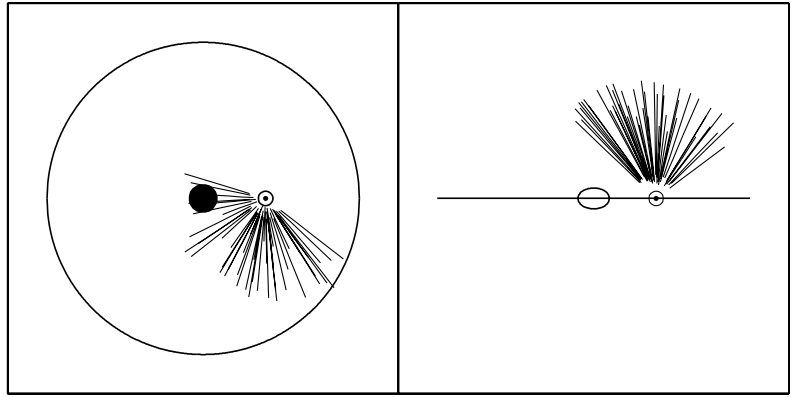}
\figcaption[Morrison.f13.eps]{Face-on (left) and edge-on views of the
Galaxy with the lines of sight traced by the 46 BTC fields shown. The
position of the Sun, galactic center and edge of disk at R=20 kpc are
shown.
\label{cartoon}}
\end{figure}

\begin{figure}
\plotone{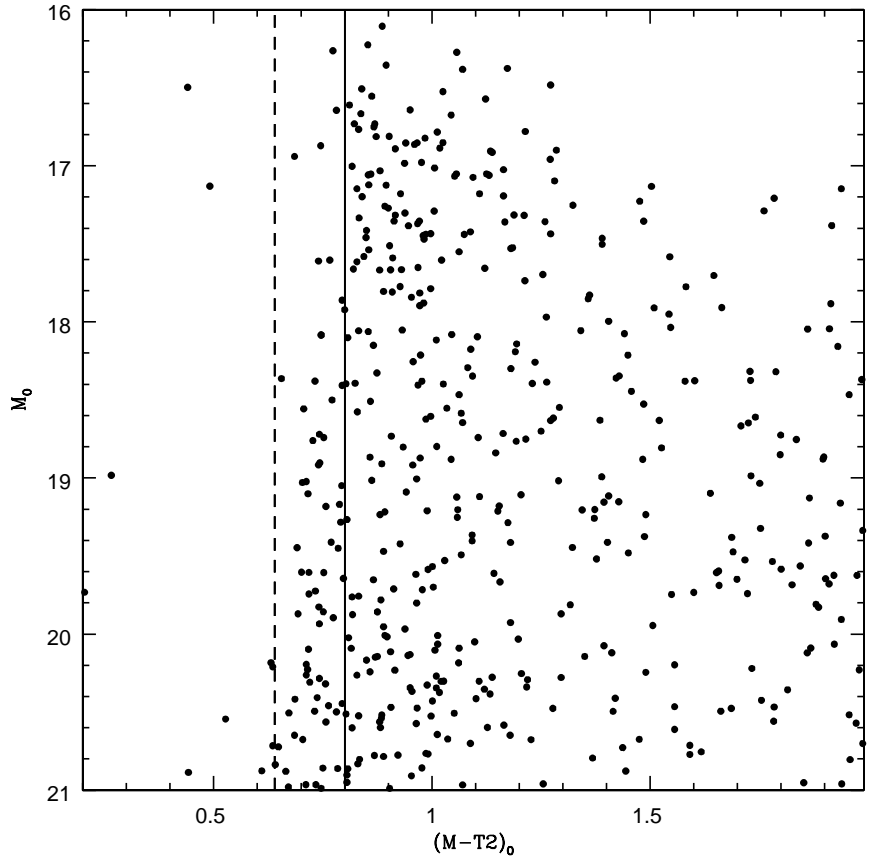}
\figcaption[Morrison.f14.eps]{Color-magnitude diagram of a field at
l=234, b=+32, with position of halo turnoff (dashed line) and thick
disk turnoff (solid line) shown. The thick disk stars are clearly
visible redward of the solid line for M brighter than 18. Halo turnoff
stars are rare for these brightnesses and become increasingly common
for fainter values of M.
\label{turnoffs}}
\end{figure}
\clearpage
\begin{figure}

\includegraphics[scale=0.6]{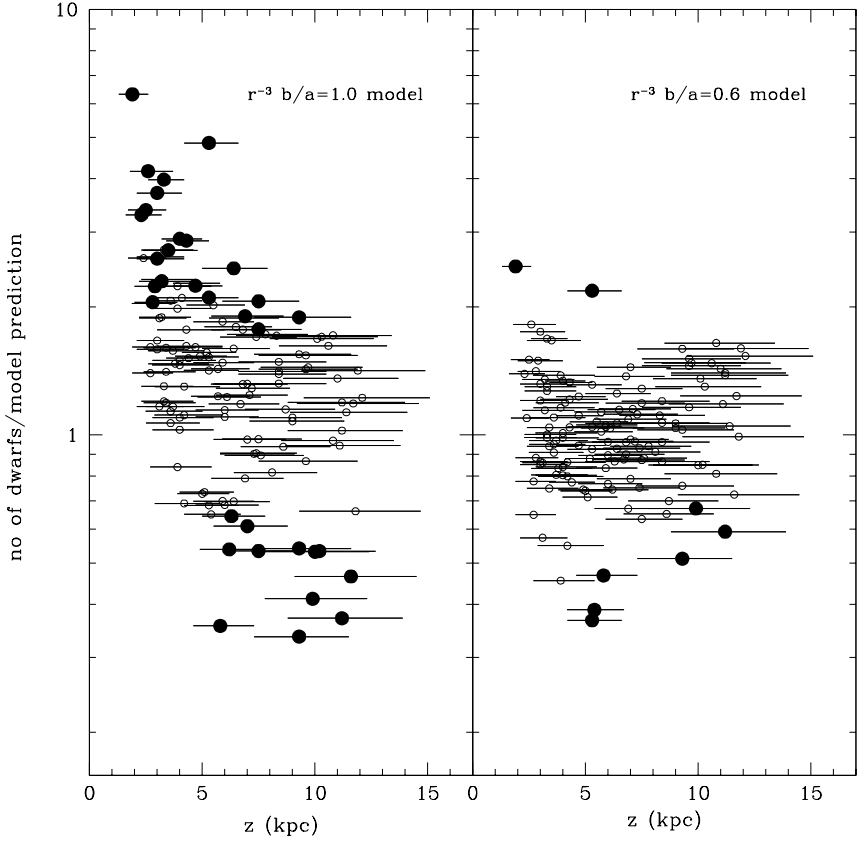}

\figcaption[Morrison.f15.eps]{Ratio of number of turnoff stars to
model prediction versus the range of z heights traversed. Points which
are more than 2.5$\sigma$ from the model are shown as solid
symbols. The local halo density was adjusted to give the best fit for
these fields: the spherical halo model uses a local density 40\% of
that given in Table \ref{halolf} and the halo with b/a=0.6 has a local
density 10\% higher than that of Table \ref{halolf}.
\label{axrat}}
\end{figure}
\clearpage
\begin{figure}
\includegraphics[scale=0.6]{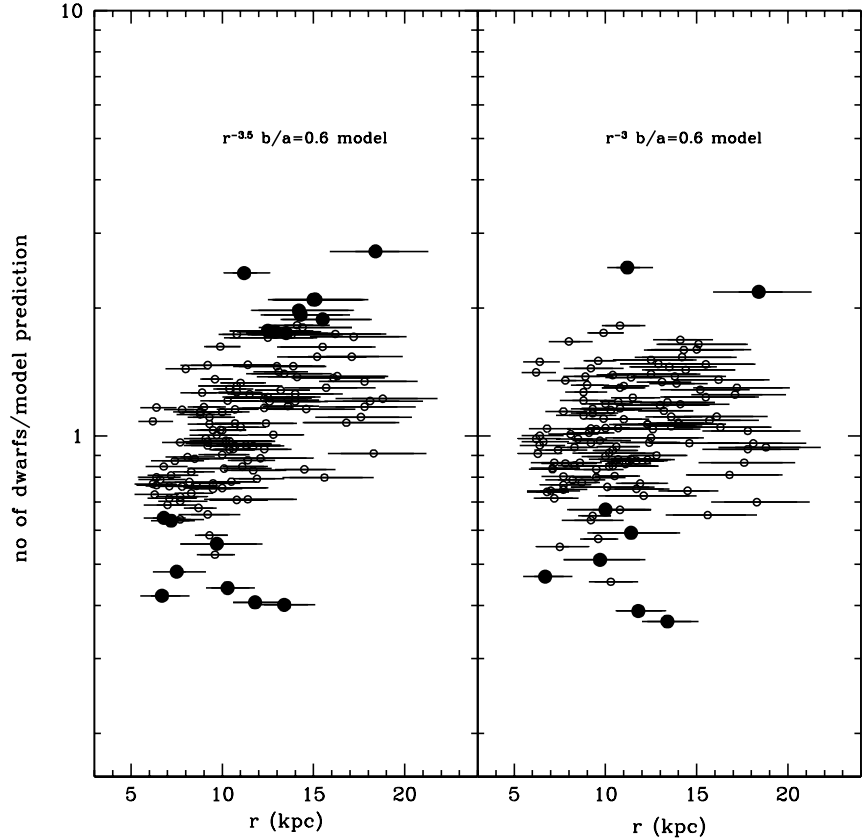}
\figcaption[Morrison.f16.eps]{Ratio of number of turnoff stars to
 model prediction versus the range of $R_{gc}$ values traversed by the
 line of sight. Points which are more than 2.5$\sigma$ from the model
 are shown as solid symbols. The local halo density was adjusted to
 give the best fit for these fields: the model with exponent --3.5
 uses a local density 40\% larger than that given in Table
 \ref{halolf} while the halo with exponent --3.0 has a local density
 10\% higher than that of Table \ref{halolf}.
\label{expon}}
\end{figure}

\clearpage
\begin{figure}
\plotone{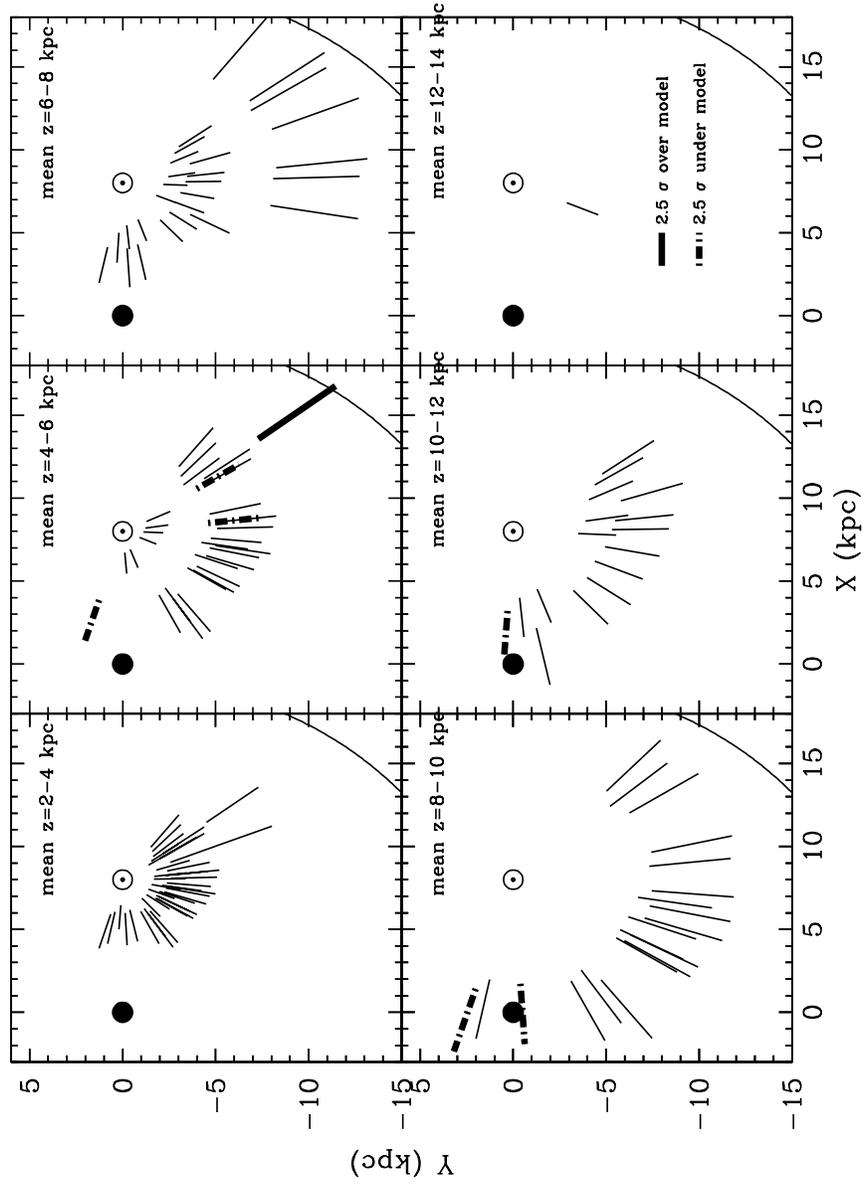}
\epsscale{0.8}
\figcaption[Morrison.f17.eps]{Face-on views of the galaxy for
different ranges of z height. Lines of sight whose halo turnoff star
numbers agree with our model are shown as solid lines. Lines of sight
which are more than 2.5$\sigma$ above the model are shown as bold
lines, and those which are more than 2.5$\sigma$ below are shown as
dashed lines. 
\label{residr}}
\end{figure}

\begin{figure}
\plotone{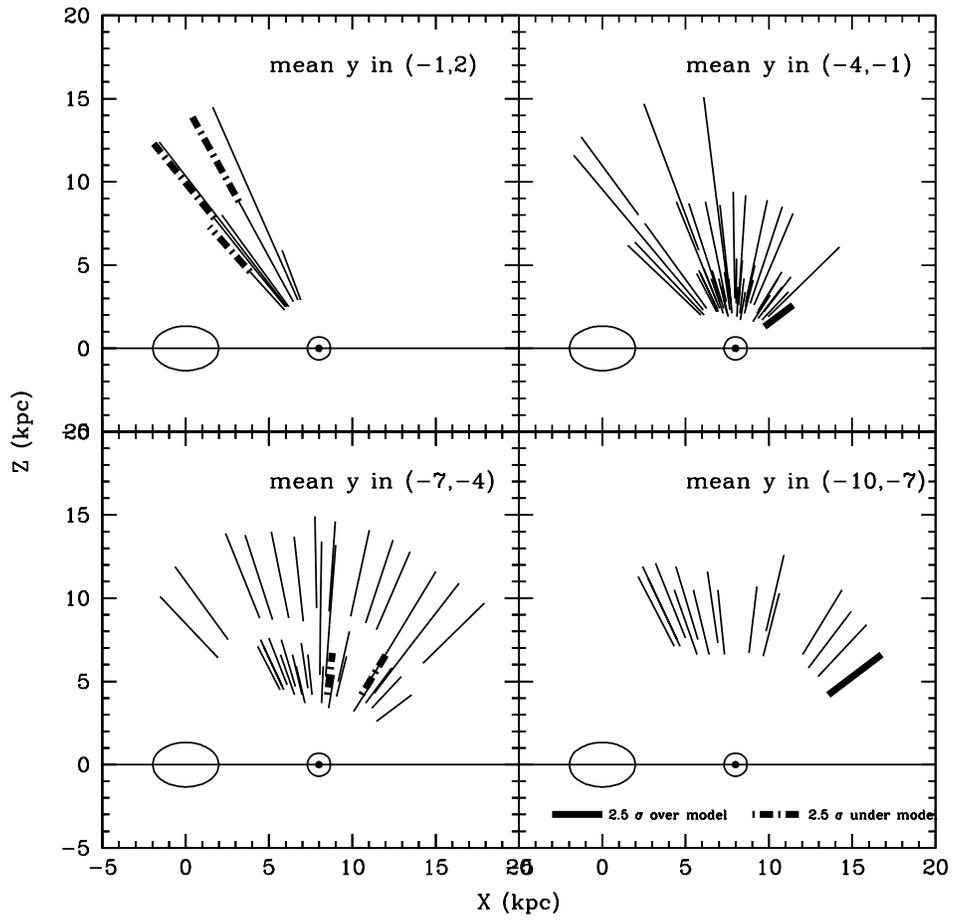}
\figcaption[Morrison.f18.eps]{Edge-on view of the lines of sight of the 46 BTC
fields. Symbols have the same meaning as in Fig.\ref{residr}.
\label{residz}}
\end{figure}

\begin{figure}
\plotone{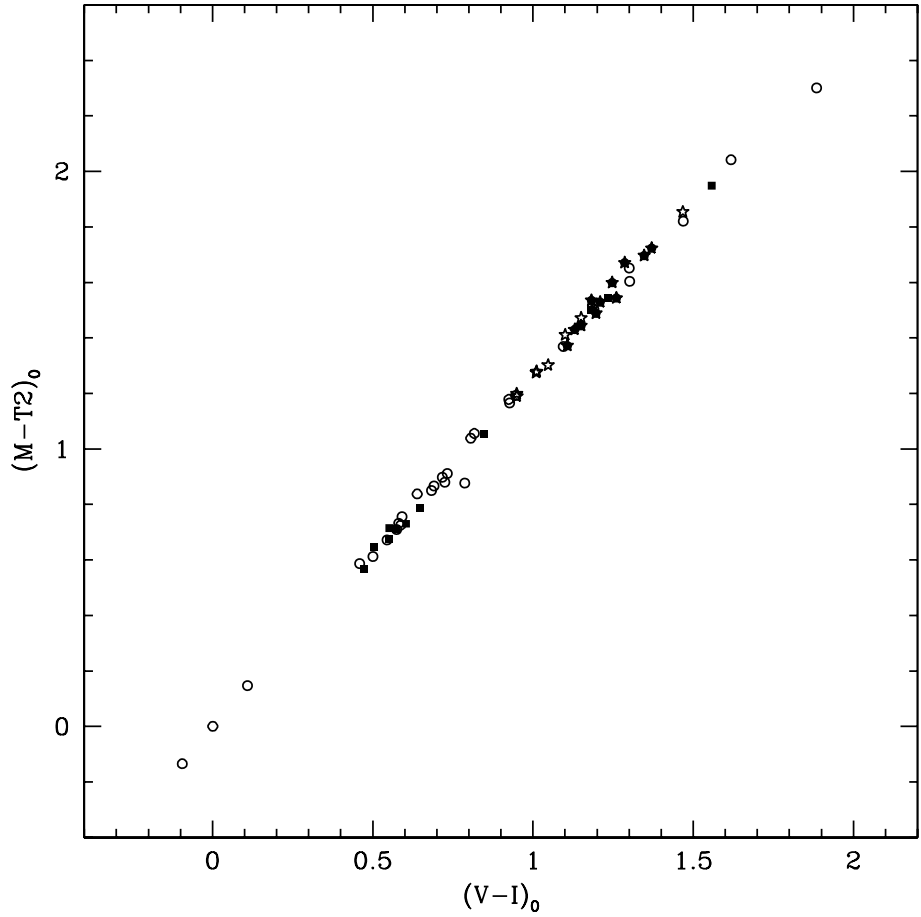}
\figcaption[Morrison.f19.eps]{Relation between \mi\ and \vi\ for
Landolt/Washington standards (giants plotted as stars, dwarfs as
filled squares, unknown spectral types as open circles) and globular cluster
giants (filled stars).  It can be seen that there is a simple
relation between Washington \mi\ color and \vi\, and that the relation
does not depend on metallicity. \label{mt2vi}}
\end{figure}

\begin{figure}
\plotone{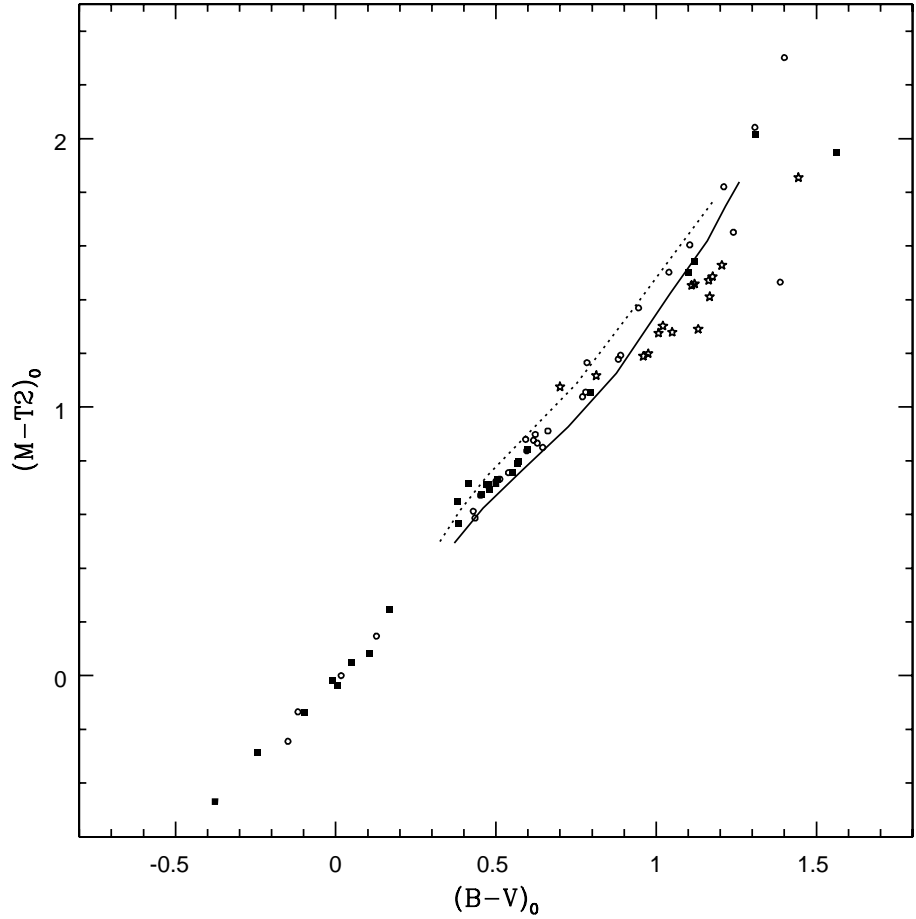}
\figcaption[Morrison.f20.eps]{Relation between Washington \mi\ and \bv\, for
Landolt/Washington standards (giants plotted as stars, dwarfs as
filled squares, unknown spectral types as open circles)and dwarfs from
\protect\citet{gpiche92}. The models of \protect\citet{pb94} for solar abundance dwarfs
(solid line) and [Fe/H] = --2.0 dwarfs (dotted line) are also shown.
\label{mt2bv}}
\end{figure}

\begin{figure}
\plotone{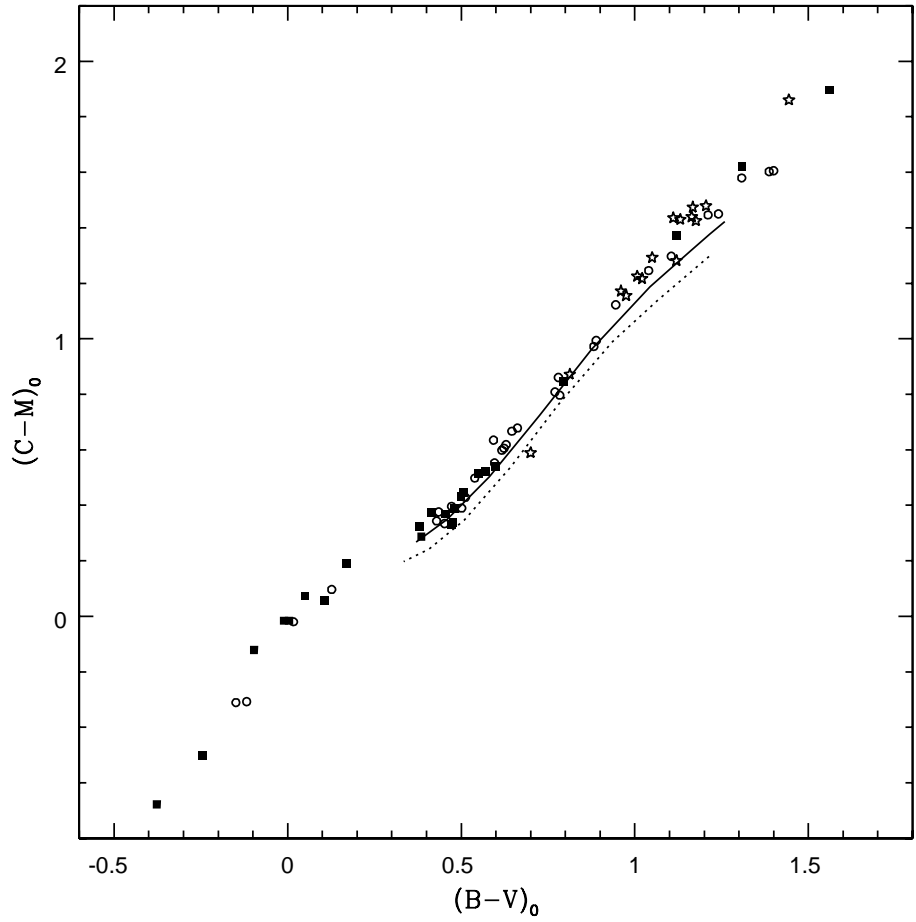}
\figcaption[Morrison.f21.eps]{Relation between Washington \cm\ and \bv\, for
Landolt/Washington standards (symbols as in Fig. \ref{mt2bv}) and
dwarfs from \protect\citet{gpiche92}. Models for dwarfs of solar metallicity
(solid line) and [Fe/H] = --2.0 (dotted line) from \protect\citet{pb94} are
also shown.
\label{cmbv}}
\end{figure}

\begin{figure}
\plotone{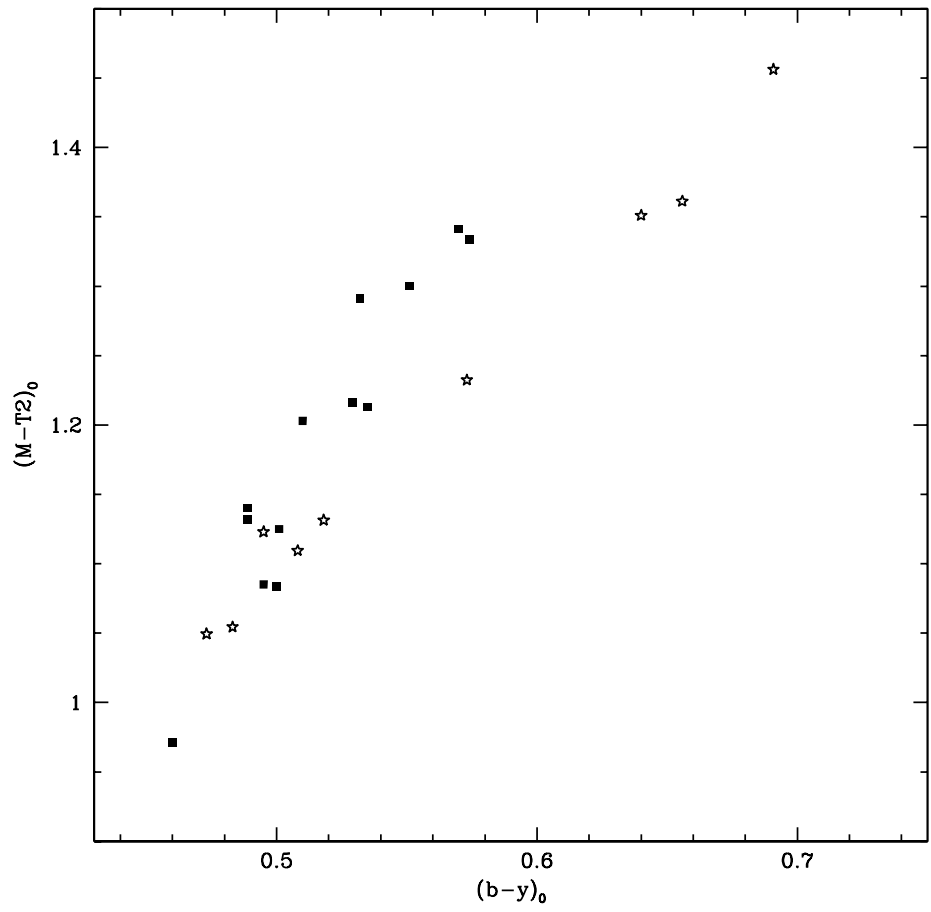}
\figcaption[Morrison.f22.eps]{Relation between Washington \mi and
Stromgren \by\ for giants and dwarfs (symbols as in Fig. \ref{mt2bv}).
\label{bymt2}}
\end{figure}

\newpage

\begin{deluxetable}{ll}
\tablewidth{0pt}
\tablecaption{Literature Sources for Fig \ref{disthist}}
\tablehead{
\colhead{Star type}           & \colhead{References}
}
\startdata
RR Lyrae & \citet{mrsh84} \\
& \citet{ciard89}\\
& \citet{nick91}\\
& \citet{wett96}\\
BHB  & \citet{nh91}\\
& \citet{cmf95} \\
& \citet{edo99} \\
Red giant & \citet{rat89}\\
& \citet{crosw91} \\
& \citet{edo99} \\
Main sequence  & \citet{srm92} \\
Carbon  & \citet{tot98} \\
\enddata
\label{distrefs}
\end{deluxetable}

\begin{deluxetable}{lclccll}
\tabletypesize{\scriptsize}
\tablewidth{0pt}
\tablecaption{Literature sources for metallicity and color for stars
in Figs. \ref{Mgb/H}}
\tablehead{
\colhead{Star ID} & \colhead{[Fe/H]}  &
\colhead{$(b\!-\!y)_0$} & \colhead{$(B\!-\!V)_0$} & \colhead{(\vi)$_0$} & \colhead{(\mi)$_0$} & \colhead{Source}
}
\startdata
HD 98281 & --0.25 &  0.46 & \nodata&\nodata & 0.97$^a$ & \citet{egg98}\\
HD 108564 & --0.52 & 0.57 &\nodata &\nodata& 1.34$^a$ & \citet{egg98} \\
HD 134440 & --1.52 & \nodata &\nodata&0.973 & 1.22$^b$ & \citet{ruth81,msb90} \\
HD 161848 & --0.18  & 0.49 &\nodata& \nodata & 1.12$^a$ & \citet{egg98} \\
HD 165195 & --2.1 & 0.73 & \nodata &\nodata & 1.53$^a$ & \citet{bond80} \\ 
HD 182488 & 0.08  & 0.48 & \nodata &\nodata& 1.07$^a$ & \citet{egg98} \\
HD 190404 & --0.44 &\nodata &\nodata &\nodata & 1.114 & \citet{doug84,egg98} \\
BD+52 1601 & --1.8 & 0.55 &\nodata &\nodata& 1.18$^a$ & \citet{bond80} \\
BD+41 3306 & --0.87 &  \nodata & 0.81&\nodata & 1.10: & \citet{ruth81,carn79}\\
BD+09 2574 & --2.4 & 0.54 & \nodata &\nodata& 1.17$^a$ & \citet{bond80,tat95} \\
BD+01 2916 & --2.0 &\nodata  & \nodata &\nodata & 1.68 & \citet{bond80,doug86} \\
BD$-$00 4234 & --0.99 & 0.58 &\nodata &\nodata& 1.36$^a$ & \citet{ruth81,tat95} \\
&&&&&&\\
G202-25 & --0.38 & \nodata & 0.87&\nodata & 1.15: & \citet{lcl88} \\
&&&&&&\\
NGC 5053 D & --2.58 & \nodata&  1.01 &\nodata& 1.3: & \citet{zinn,sand77} \\
&&&&&&\\
M3 Cud205 & --1.66 &\nodata & 1.36 &\nodata& 1.75: & \citet{zinn,cud79} \\
M3 Cud250 & --1.66 & \nodata& 0.94 &\nodata& 1.20: & \\
M3 Cud354 & --1.66 & \nodata & 0.82&\nodata & 1.10: & \\
M3 Cud1327 & --1.66  & \nodata & 0.74 &\nodata & 1.0: & \\
&&&&&&\\
NGC 6171 16 & --0.99 &\nodata & 1.11 &\nodata& 1.45: &  \citet{zinn,sand64} \\
NGC 6171 20  & --0.99 &\nodata &  1.06& \nodata& 1.3: &  \\
NGC 6171 62 & --0.99 & \nodata& 1.29 &\nodata& 1.7:& \\
&&&&&&\\
M71 a1 & --0.58 &\nodata & 1.30 &\nodata& 1.7: & \citet{zinn,cud85} \\
M71 l1 & --0.58 &\nodata & 0.99 &\nodata &1.3: & \\
M71 s232 & --0.58 &\nodata & 1.18 &\nodata &1.5: & \\
 &&&&&&\\
M2 2 & --1.62 &\nodata &\nodata &1.20 & 1.50 & \citet{zinn,taft91} \\
\enddata
\tablenotetext{a}{These values of \mi were derived from \by.}
\tablenotetext{b}{These values of \mi were derived from \vi.}
\tablenotetext{:}{These values of \mi were derived from \bv and
are likely to be less accurate than those derived from \by or \vi.}
 
\label{lumrefs}
\end{deluxetable}

\begin{deluxetable}{llll}
\tablewidth{0pt}
\tablecaption{Turnoff colors for Globular Clusters}
\tablehead{
\colhead{Cluster}   & $(b-y)_0$ & (\vi)$_0$          & \colhead{References}
}
\startdata
M92 & \nodata & 0.531 & \citet{jj98} \\
NGC 7099  & \nodata &0.485 & \citet{sandq}\\
NGC 6397 &0.294 & \nodata & \citet{atts}\\
47 Tucanae & \nodata & 0.70 & \citet{kaluz}\\
\enddata
\label{turnoffcolors}
\end{deluxetable}

\begin{deluxetable}{ccc}
\tablewidth{0pt}
\tablecaption{Halo turnoff star luminosity function for stars with \bv\
between 0.38 and 0.50}
\tablehead{
\colhead{$M_V$} & \colhead{$\Phi$ (stars per kpc$^3$)}  &
\colhead{Number of stars in CLL sample}
}
\startdata
3.5 & 73.1 & 2 \\
4.0 & 986.0 & 13 \\
4.5 & 2695.7 & 46 \\
5.0 & 807.8 & 17 \\
5.5 & 151.2 & 5 \\
\enddata
\label{halolf}
\end{deluxetable}


\newpage
 
\begin{thebibliography}{dum}
\bibitem[Anthony-Twarog et al.(1992)]{atts} Anthony-Twarog, B.J., Twarog,
B.A. \& Suntzeff, N.B. 1992, AJ, 103, 1264
\bibitem[Armandroff and Da Costa(1991)]{taft91} Armandroff, T.E. \& Da
Costa, G.S. 1991, AJ, 101, 1329
\bibitem[Bahcall and Soniera(1984)]{bsmodel} Bahcall, J.N. \& Soniera,
R.M. 1984, ApJS, 55, 67
\bibitem[Bahcall and Casertano(1986)]{bc86} Bahcall, J.N. \& Casertano,
S. 1986, ApJ, 308, 347
\bibitem[Barbuy et al.(1992)]{barb92} Barbuy, B., Erdelyi-Mendes,
M. \& Milone, A. 1992, A\&AS 93, 235
\bibitem[Beers et al.(1985)]{bps85} Beers, T.C., Preston, G.W. \& Shectman,
S.A. 1985, AJ, 90, 2089
\bibitem[Bell et al.(1994)]{bell94} Bell, R.A., Paltoglou, G. \&
Trippico, M.A. 1994, MNRAS, 268, 771
\bibitem[Bessell(1979)]{msb79} Bessell, M.S. 1979, PASP, 91, 589
\bibitem[Bessell(1990)]{msb90} Bessell, M.S. 1990, A\&AS 83, 357
\bibitem[Bidelman and MacConnell(1973)]{bm73} Bidelman, W.P. \&
MacConnell, D.J. 1973, AJ, 78, 687
\bibitem[Bond(1980)]{bond80} Bond, H.E. 1980, ApJS, 44, 517
\bibitem[Canterna(1976)]{can76} Canterna, R.W. 1976, AJ, 81, 228
\bibitem[Carbon et al.(1987)]{carbon} Carbon, D.F., Barbuy, B., Kraft,
R.P., Friel, E.D. \& Suntzeff, N.B. 1987, PASP, 99, 335
\bibitem[Carney(1979)]{carn79} Carney, B.W. 1979, ApJ, 233, 211
\bibitem[Carney and Latham(1987)]{cl87} Carney, B.W. \& Latham, D.W. 1987,
AJ, 93, 116
\bibitem[Carney et al.(1989)]{carn89} Carney, B.W., Latham, D.W. \& Laird,
J.B. 1989, AJ, 97, 423
\bibitem[Carney et al.(1994)]{carn94} Carney, B.W., Latham, D.W., Laird,
J.B. \& Aguilar, L.A. 1994, AJ, 107, 2240
\bibitem[Ciardullo et al.(1989)]{ciard89} Ciardullo, R., Jacoby, G.H. \&
Bond, H.E. 1989, AJ, 98, 1649
\bibitem[C\^ot\'e et al.(1993)]{cote} C\^ot\'e, P., Welch, D.L., Fischer,
P., \& Irwin, M.J. 1993, ApJ, 406, L59
\bibitem[Croswell et al.(1991)]{crosw91} Croswell, K., Latham, D.W.,
Carney, B.W., Schuster, W. \& Aguilar, L.A. 1991, AJ, 101, 2078
\bibitem[Cudworth(1979)]{cud79} Cudworth, K.M. 1979, AJ, 84, 1312
\bibitem[Cudworth(1985)]{cud85} Cudworth, K.M. 1985, AJ, 90, 65
\bibitem[Da Costa and Armandroff(1990)]{garytaft} Da Costa, G.S. \&
Armandroff, T.E. 1990, AJ, 100, 162
\bibitem[Davis et al.(1985)]{davis85} Davis, M.M. et al 1985, \apj, 292,
371
\bibitem[Dohm-Palmer et al.(1999)]{robbie} Dohm-Palmer, R.C., Mateo, M.L.,
Harding, P., Olszewski, E.W., Morrison, H.L., Freeman, K.C. \& Norris,
J.E. 1999, in preparation.
\bibitem[Eggen et al.(1962)]{egg62} Eggen, O. J., Lynden-Bell, D., \&
Sandage, A. R. 1962, \apj, 136, 748
\bibitem[Eggen(1998)]{egg98} Eggen, O.J. 1998, AJ, 115, 2397
\bibitem[Fleming et al.(1986)]{tom86} Fleming, T.A., Liebert, J. \&
Green, R.F. 1986, ApJ, 308, 176
\bibitem[Flynn and Morrison(1990)]{fm90} Flynn, C. and Morrison,
H.L. 1990, AJ, 100, 1181
\bibitem[Flynn et al.(1995)]{cmf95} Flynn, C., Sommer-Larsen, J.,
Christensen, P.R. \& Hawkins, M.R.S. 1995, A\&AS 109, 171
\bibitem[Geisler(1984)]{doug84} Geisler, D. 1984, PASP, 96, 723
\bibitem[Geisler(1986)]{doug86} Geisler, D. 1986, PASP, 98, 762
\bibitem[Geisler(1990)]{doug90} Geisler, D. 1990, PASP, 102, 344
\bibitem[Geisler(1995)]{doug95} Geisler, D. 1995, private communication
\bibitem[Geisler(1996)]{doug96} Geisler, D. 1996, AJ, 111, 480
\bibitem[Geisler et al.(1991)]{doug91}  Geisler, D., Claria, J.J. \&
Minniti, D. 1991, AJ, 102, 1836
\bibitem[Geisler et al.(1992)]{doug92} Geisler, D., Minniti, D. \&
Claria, J.J. 1992, AJ, 104, 627
\bibitem[Geisler et al.(1997)]{doug97} Geisler, D., Claria, J. \&
Minniti, D. 1991, PASP, 109, 799
\bibitem[Gonzalez and Pich\'e(1992)]{gpiche92} Gonzalez, G. \&
Pich\'e, F. 1992, AJ, 103, 2048
\bibitem[Governato et al.(1997)]{gov97} Governato, F., Moore, B., Cen,
R., Stadel, J., Lake, G. \& Quinn, T. 1997, NewA, 2, 91
\bibitem[Gunn(1995)]{gunn} Gunn, J.E. 1995, AAS 186, 44.05
\bibitem[Harding et al.(1999)]{paul99} Harding, P.,Morrison, H.L.,Mateo,
M.L., Olszewski, E.W., Freeman, K.C. \& Norris, J.  1999, in preparation.
\bibitem[Harris and Canterna (1979)]{hc79} Harris, H.C., and
Canterna, R.W. 1979, AJ, 84, 1750
\bibitem[Harris and Stetson(1988)]{hstet88} Harris, W.E. \& Stetson,
P.B. 1988, AJ, 96, 909
\bibitem[Hartwick(1987)]{hart87} Hartwick, F.D.A. 1987, in The
Galaxy, edited by G. Gilmore and B. Carswell (Reidel, Dordrecht), p281
\bibitem[Hauschildt et al.(1999a)]{haus99a} Hauschildt, P.H., Allard,
F. \& Baron, E. 1999, ApJ, 512, 377
\bibitem[Hauschildt et al.(1999b)]{haus99b} Hauschildt, P.H., Allard,
F., Ferguson, J.,  Baron, E. \& Alexander, D.  1999, ApJ in press, to
appear November 1999
\bibitem[Hawkins(1984)]{mrsh84} Hawkins, M.R.S. 1984, MNRAS, 206, 433
\bibitem[Helmi et al.(1999)]{helmi} Helmi, A., White, S.D.M., de
Zeeuw, P.T. \& Zhao, H-S. 1999, Nature, 402, 53
\bibitem[Hesser et al.(1987)]{47hesser} Hesser, J.F., Harris, W.E.,
Vandenberg, D.A., Allwright, J.W.B., Shott, P. \& Stetson, P.B. 1987, PASP,
99, 739
\bibitem[Ibata et al.(1994)]{igi} Ibata, R.A., Gilmore, G. \& Irwin,
M.J. 1994, \nat, 370, 194
\bibitem[Johnson and Bolte(1998)]{jj98} Johnson, J.A. \& Bolte, M. 1998,
AJ, 115, 693
\bibitem[Johnston et al.(1996)]{jhb} Johnston, K.V., Hernquist, L. \&
Bolte, M. 1996, ApJ, 465, 278
\bibitem[Johnston et al.(1995)]{jsh} Johnston, K.V., Spergel, D.V. \&
Hernquist, L. 1995, ApJ, 451, 598
\bibitem[Kaluzny et al.(1998)]{kaluz} Kaluzny, J. Wysoka, A., Stanek,
K.Z. \& Krzeminski, W. 1998, Acta Astronomica, 48, 439
\bibitem [Kinman et al.(1965)]{kin65} Kinman, T.D., Wirtanen, C.A., \&
Janes, K.A. 1965, ApJS, 11, 223
\bibitem[Kinman et al.(1994)]{tdk94} Kinman, T.D., Suntzeff, N.B \&
Kraft, R.P. 1994, AJ, 108, 1722
\bibitem[Kinman et al.(1996)]{tdk96} Kinman, T.D., Pier, J. 
R., Suntzeff, N. B., Harmer, D. L., Valdes, F., Hanson, R. B., Klemola, A. 
R. \& Kraft, R. P. 1996, AJ, 111, 1164 
\bibitem[Klypin et al.(1999)]{klyp99} Klypin, A., Kravtsov, A.V.,
Valenzuela, O. \& Prada, F. 1999 astro/ph 9901240
\bibitem[Kurucz(1992)]{kurucz92} Kurucz, R.L. 1992, Rev. Mexicana
Astron. Astrofis., 23, 181
\bibitem[Landoldt(1992)]{arlo} Landoldt, A. 1992, AJ, 104, 340
\bibitem[Laird et al.(1988)]{lcl88} Laird, J.B., Carney, B.W. \& Latham,
D.W. 1988, AJ, 95, 1843
\bibitem[Lynden-Bell(1999)]{lb3} Lynden-Bell, D. 1999, in ``The
Galactic Halo'', ASP conf. series 165, ed. B. Gibson, T. Axelrod and
M. Putnam
\bibitem[Lynden-Bell and Lynden-Bell (1995)]{lb2} Lynden-Bell, D. \&
Lynden-Bell, R.M. 1995, MNRAS, 275, 429
\bibitem[Majewski(1992)]{srm92} Majewski, S.R. 1992, ApJS, 78, 87
\bibitem[Majewski et al.(1994)] {srm94} Majewski, S.R., Munn, J.A., \&
Hawley, S.L. 1994, ApJ, 427, L37
\bibitem[Mateo (1998)]{mm98} Mateo, M.L. 1998 ARAA 36, 435
\bibitem[Mendez et al.(1999)]{mend99} Mendez, R.A., Platais, I.,
Girard, T.M.,  Kozhurina-Platais, V., van Altena, W.F. 1999, ApJ, in press. 
\bibitem[Morrison (1993)]{hlm93} Morrison, H.L. 1993, AJ, 106, 578
\bibitem[Morrison et al.(1990)]{mff} Morrison, H.L., Flynn, C. \& Freeman,
K.C. 1990, AJ, 100, 1191
\bibitem[Morrison et al.(2000)]{hlm00} Morrison, H.L., Mateo, M.,
Olszewski, E.W., Harding, P., Dohm-Palmer, R., Freeman, K.C., \&
Norris, J.E. 2000, in preparation 
\bibitem[Norris \& Hawkins (1991)]{nh91} Norris, J.E. and Hawkins,
M.R.S. 1991, ApJ, 380, 104
\bibitem[Olszewski et al.(1999)]{edo99} Olszewski, E.W., Harding, P.,
Mateo, M.L., Morrison, H.L., Freeman, K.C. \& Norris, J. 1999, in
preparation
\bibitem[Paltoglou and Bell (1994)]{pb94} Paltoglou, G. \& Bell, R.A. 1994
MNRAS, 268, 793
\bibitem[Pier(1982)]{pier82} Pier, J. R. 1982, \aj, 87, 1515 
\bibitem[Pier(1983)]{pier83} Pier, J.R. 1983, ApJS, 52, 791
\bibitem[Pier(1984)]{pier84} Pier, J. R. 1984, \apj, 281, 260
\bibitem[Peterson (1981)]{ruth81} Peterson, R.C. 1981, ApJ, 244, 989
\bibitem[Preston et al.(1994)]{pbs94} Preston, G.W., Beers, T.C. \&
Shectman, S.A. 1994, AJ, 108, 538
\bibitem[Preston et al.(1991)]{psb91} Preston, G.W., Shectman, S.A., and
Beers, T.C. 1991, ApJ, 375, 121
\bibitem[Ratnatunga and Freeman (1989)]{rat89} Ratnatunga, K.U. \&
Freeman, K.C. 1989, ApJ, 339, 126
\bibitem[Reid \& Majewski (1993)]{inr93} Reid, N., \& Majewski, S.R. 1993,
pJ, 409, 635
\bibitem[Ryan and Norris (1991)]{sean91} Ryan, S.G. and Norris, J.E. 1991,
AJ, 101, 1835
\bibitem[Saha (1985)]{abi} Saha, A. 1985, ApJ, 289, 310
\bibitem[Sandage and Katem (1964)]{sand64} Sandage, A. \& Katem, B. 1964,
ApJ, 139, 1088
\bibitem[Sandage et al.(1977)]{sand77} Sandage, A., Katem, B. \& Johnson,
H.L. 1977, AJ, 82, 389
\bibitem[Sandage and Fouts (1987)]{sf87} Sandage, A. \& Fouts, G. 1987,
AJ, 93, 74
\bibitem[Sandage and Luyten(1969)]{sand69} Sandage, A. \& Luyten,
W.J. 1969, ApJ, 155, 913
\bibitem[Sandquist et al.(1999)]{sandq} Sandquist, E.L., Bolte, M.,
Langer, G.E., Hesser, J.E \& Mendes de Olivera, C 1999, ApJ, 518, 262
\bibitem[Seitter(1970)]{bonner} Seitter, W.C. 1970, Atlas for
Objective Prism Spectra, Bonner Spektral Atlas I,  Ferd. Dummlers Verlag,
Bonn
\bibitem[Schlegel et al.(1998)]{schlegel} Schlegel, D.J., Finkbeiner,
D.P. \& Davis, M. 1998, ApJ, 500, 525
\bibitem[Searle and Zinn(1978)]{szinn} Searle, L. \& Zinn, R. 1978, ApJ,
225, 357
\bibitem[Sommer-Larsen and Christensen (1985)]{slc} Sommer-Larsen, J. \&
Christensen, P.R. 1985, MNRAS, 212, 851
\bibitem[Suntzeff et al.(1991)]{nick91} Suntzeff, N.B., Kinman, T.D. and
Kraft, R.P. 1991, ApJ, 367, 528
\bibitem[Totten and Irwin (1998)]{tot98} Totten, E.J. \& Irwin, M.J. 1998,
MNRAS, 294, 1
\bibitem[Twarog and Anthony-Twarog (1995)]{tat95} Twarog, B.A. \&
Anthony-Twarog, B.J. 1995, AJ, 109, 2828
\bibitem[Unavane et al.(1996)]{unavane} Unavane, M., Wyse, R.F.G. \&
Gilmore, G. 1996, MNRAS, 278, 727
\bibitem[van Altena et al.(1991)]{yale} van Altena, W.M.F., Lee, J. \&
Hoffleit, E. 1991, The General Catalog of Trigonometric Parallaxes, a
Preliminary Version (Yale University Observatory, New Haven)
\bibitem[Wetterer and McGraw (1996)]{wett96} Wetterer, C.J. \& McGraw,
J.T. 1996, AJ, 112, 1046
\bibitem[Wyse and Gilmore(1989)]{wg89}  Wyse, R.F.G. \& Gilmore, G. 1989,
Comments in Astrophys., 13, 135
\bibitem[Zinn(1985)]{zinn} Zinn, R.J. 1985, ApJ, 293, 424
\bibitem[Zinn(1993)]{zinn93} Zinn, R.J., 1993, in {\it The Globular
Cluster- Galaxy Connection,} ASP Conf Series 48, edited by G.~H.~Smith
\& J.~P.~Brodie, (ASP, San Francisco), p.~38
\end{thebibliography}
\end{document}